\documentclass{aa}  

\usepackage{graphicx}
\usepackage{txfonts}
\usepackage{enumitem}
\usepackage{adjustbox}
\usepackage{float}
\usepackage{hyperref}
\usepackage{xcolor, soul}
\hypersetup{
    colorlinks=true,
    linkcolor=black,
    filecolor=magenta,      
    urlcolor=cyan,
    citecolor=blue
    }
    
\usepackage[switch]{lineno} 

\begin{document}

   \title{Investigating the Growth of Little Red Dot Descendants at $z<4$ with the JWST}

\author{Jean-Baptiste Billand \inst{1}
    \and David Elbaz\inst{1}
    \and Fabrizio Gentile\inst{1,2}
    \and Maxime Tarrasse\inst{1}
    \and Maximilien Franco\inst{1}
    \and Benjamin Magnelli\inst{1} 
    \and Emanuele Daddi\inst{1}
    \and Yipeng Lyu \inst{1}
    \and Avishai Dekel \inst{3}
    \and Fabio Pacucci \inst{4,5}
    \and Valentina Sangalli \inst{1}
    \and Mark Dickinson\inst{6}
    \and Mauro Giavalisco \inst{7}
    \and Benne W. Holwerda \inst{8}  
    \and Dale D. Kocevski \inst{9}
    \and Anton M. Koekemoer\inst{10}
    \and Vasily Kokorev \inst{11}
    \and Ray A. Lucas \inst{10}
    \and Pablo G. Pérez-González \inst{12}
          }

\institute{
Université Paris-Saclay, Université Paris Cité, CEA, CNRS, AIM,  91191, Gif-sur-Yvette, France
\and INAF – Osservatorio di Astrofisica e Scienza dello Spazio, via Gobetti 93/3 - 40129, Bologna - Italy
\and Racah Institute of Physics, The Hebrew University, Jerusalem 91904 Israel
\and Center for Astrophysics | Harvard \& Smithsonian, 60 Garden St, Cambridge, MA 02138, USA
\and Black Hole Initiative, Harvard University, 20 Garden St, Cambridge, MA 02138, USA 
\and NSF’s National Optical-Infrared Astronomy Research Laboratory, 950 North Cherry Avenue, Tucson, AZ 85719, USA
\and University of Massachusetts Amherst, 710 North Pleasant Street, Amherst, MA 01003-9305, USA
\and Department of Physics and Astronomy, University of Louisville, Natural Science Building 102, 40292 KY Louisville, USA
\and Department of Physics and Astronomy, Colby College, Waterville, ME 04901, USA
\and Space Telescope Science Institute, 3700 San Martin Drive, Baltimore, MD 21218, USA
\and Department of Astronomy, The University of Texas at Austin, Austin, TX 78712, USA 
\and Centro de Astrobiología (CAB), CSIC-INTA, Ctra. de Ajalvir km 4, Torrejón de Ardoz, E-28850, Madrid, Spain
}

   \date{}

  \abstract{One of the most remarkable and unexpected results of the James Webb Space Telescope is the discovery of a population of compact red galaxies: the so-called Little Red Dots (LRDs). The existence of these galaxies raises many questions, including that of their nature and origin, as well as that of their evolution. Indeed, these compact red sources exhibit a pronounced decline in number density by nearly two orders of magnitude from \( z = 6 \) to \( z = 3 \).}
  {In this paper, we investigate the possible evolution of this galaxy population at lower redshift. To this end, we have identified a sample of candidates in the CEERS images that could represent the descendants of LRDs by assuming a single evolutionary path: the development of a blue star-forming outskirt while retaining a red inner core. }
  {Our color–magnitude selection identifies red galaxies as red as LRDs at \( z < 4 \), defined by a compact red inner region and blue outskirts. The red core is associated with the LRD population, while the blue periphery traces recently formed young stars. Morphological properties were derived by fitting single Sérsic profiles, while other physical quantities were obtained through spectral energy distribution fitting, assuming a stellar-only model for both the inner region and the outskirts.
}
  {The selected galaxies are likely "Post-LRDs" galaxies, showing similar properties to LRDs under a stellar-only model: stellar masses of $M_* \approx 10^{10} \, M_\odot$, central densities $\Sigma_* \approx 10^{11} \, M_\odot \, \text{kpc}^{-2}$, similar rest-frame red colors, and a $\sim$1 kpc offset below the size–mass relation. Their number density at $z=3 \pm 0.5$ ($ 10^{-4.15} \, \text{Mpc}^{-3}$) matches that of LRDs at $5 < z < 7$, supporting an evolutionary connection. We find a strong redshift-dependent increase in both outskirts mass fraction and galaxy size-from $\sim$250 pc at $z=5$ to $\sim$600 pc at $z=3$-indicating overall stellar growth. Meanwhile, the core remains as red and as massive, but the characteristic V-shaped SED fades as the extended star-forming envelope becomes dominant.

}
  { These findings support an evolutionary scenario in which LRDs gradually acquire an extended stellar component over cosmic time by cold accretion. This growth affects the initial LRD state in two key ways: the physical size increases, and the characteristic V-shaped SED in the core becomes less distinct and disappears. As a result, the original selection criteria based on both of them can no longer identify this population as it evolves, providing an explanation for their observed decline in number density.
}


\authorrunning{name(s) of author(s)}
   \maketitle

%

\section{Introduction}

The James Webb Space Telescope (JWST) has enabled the discovery of a particularly enigmatic class of galaxies: a large population of red and compact objects referred to as Little Red Dots \citep[LRDs,][]{matthee_little_2024, kocevski_hidden_2023, kocevski_rise_2024, labbe_uncover_2023}. Initially presumed to be purely Active Galactic Nuclei (AGN) due to their broad Balmer lines,  with FWHM$\approx2000-4000$ km s$^{-1}$, associated with the broad line region (BLR) \citep{kocevski_hidden_2023, greene_uncover_2024, kocevski_rise_2024, Taylor_broad_line_agn_2024, wang_rubies_2024}, hints of a non-negligible stellar component in the observed flux began to challenge our understanding of these sources. Indeed, the spectral energy distribution (SED) flattens in the rest-frame near infrared, and a stellar component has been suggested to better explain the rest-frame observed SED instead of a pure AGN torus \citep{williams_galaxies_2024,   perez-gonzalez_what_2024}. Several definitions of LRDs have been employed since then, each focusing on red and compact objects. In the literature, they can be selected based on their broad Balmer lines \citep{kokorev_uncover_2023, kocevski_hidden_2023, matthee_little_2024, furtak_high_2024} or through color-color selections. Among these, the criterion that selects a rest-frame UV excess and a red optical slope, forming the V-shape is used \citep{kocevski_rise_2024, leung_exploring_2024, euclid_LRDs}, as well as more direct color-color diagrams that identify objects with red and blue colors \citep{labbe_population_2023, kokorev_census_2024, akins_cosmos-web_2024, barro_photometric_2024, xiao_nocII_2025, Ma_counting_LRD_2025}. All selection criteria share a common requirement: the compactness of the galaxies, typically defined in the observed \( 4.4 \, \mu\text{m} \) band, as follows or in a similar manner: \( F_\text{F444W}(0.4'')/F_\text{F444W}(0.2'') < 1.7 \) \citep{labbe_uncover_2023, akins_cosmos-web_2024, kokorev_census_2024, greene_uncover_2024, barro_photometric_2024, xiao_nocII_2025}. The half-light radius can also serve as a compactness criterion: \( R_e < 1.5 \, r_\text{h,star} \), where \( r_\text{h,star} \) is the effective half-light radius of stars \citep{kocevski_rise_2024}.

When the V-shape defined on the UV and optical slope by \citet{kocevski_rise_2024} is observed, the inflection point appears to occur at the same rest-frame wavelength as the Balmer break, which is known to be associated with old stellar populations \citep{setton_little_2024}. Moreover, if LRDs are assumed to be pure AGN, the accretion rate is expected to vary over time, leading to variable flux \citep{Ulrich_variability_1997, macleod_a_description_2012}, which has not yet been observed in the UV or X-ray bands \citep{kokubo_challenging_2024, zhang_analysis_2024, tee_lack_of_rest}. Additionally, the broadening of the lines can also be explained by the compact kinematics of a star-forming galaxy \citep{perez-gonzalez_what_2024, akins_cosmos-web_2024, Baggen_the_small_2024}. However, the first instance of time variability in an LRD was detected in A2744-QSO1 \citep{furtak_investigating_2025} by measuring the equivalent width (EW) of $H_\alpha$ and $H_\beta$. Approximately $20\%$ variability was observed over a 2.4-year period, confirming that the flux is dominated by an AGN, at least for this particular LRD. On one hand, \citet{xiao_nocII_2025} attempted to detect [CII] lines to differentiate between an AGN and dusty star-formation for two V-shaped LRDs at $z_{\text{spec}} > 7$. The non-detection of [CII] disfavors the only-stellar model. This is further supported by the lack of detection in sub-millimeter observations with ALMA at 1.2 mm \citep{akins_cosmos-web_2024, casey_dust_2024, labbe_uncover_2023}, as well as by the upper limit on the dust mass of $10^6 \, M_{\odot}$ estimated by \citet{casey2025upperlimit106modot} in a sample of 60 LRDs.  \\

Due to the variety of definitions, not all LRDs exhibit a clear V-shape, broad lines, or a Balmer break. Approximately $60\%$ of LRDs display broad lines \citep{kocevski_rise_2024, greene_uncover_2024, Leob2024}, and $>70\%$ of V-shaped LRDs show this feature, suggesting a heterogeneous population governed by diverse physical processes. The degeneracy between the two interpretation --  pure AGN and pure stellar model-- is well described in \citet{akins_cosmos-web_2024}, they studied 434 LRDs as part of the COSMOS-Web survey \citep{Cosmos_survey} between $z \approx 5-9$. If they are interpreted as purely AGN, the inferred accretion rate from the LRDs alone competes with that of all AGN at cosmic noon, i.e. $z\approx2$, approximately $10^{-4} { M}_{\odot}  \text{ yr}^{-1}\text{ Mpc}^{-3}$ \citep{Yang_BH_accretion_2023}. The black hole mass inferred would also be overmassive with respect to the stellar host component \citep{Pacucci2023,Maiolino_2023, Durodola_2025}. If they are interpreted as purely stellar, inferred stellar masses are very important, above $10^{10}M_\odot$ and JWST/MIRI is key to inferred robust masses consistent with $\Lambda$CDM. The inferred density profile is broadly in agreement with the densest core observed in the universe up to $z \approx 3 $ by \citet{hopkins2010}, though slightly higher. The stellar density can even reach the threshold required for runaway stellar collisions to occur \citet{Guia2024}.

In both cases, the physical properties of LRDs reach extreme values, suggesting that a combination of both models might be necessary. Significant Balmer break strengths have been observed in several LRDs \citep{Naidu_bhstar, degraaff_a_remarkable, Taylor_2025_capers}, approximately four times higher than those in massive quiescent galaxies. This strength challenges the stellar origin of the Balmer break in LRDs. As the usual point of view of AGN is also challenged by previous studies, a new explanation of LRDs has been suggested. In this model, an LRD is interpreted as an AGN surrounded by extremely dense gas (\(n_{\text{H}} \approx 10^{11} \ \text{cm}^{-3}\), \(N_{\text{H}} \approx 10^{25} \ \text{cm}^{-2}\)), a scenario first proposed by \citet{Ji_black_thunder, Inayoshi_2025} to account for the \(\mathrm{H}\beta\) absorption lines observed in several LRDs. This configuration also enables the presence of an extremely strong Balmer break, as developed in \citet{Ji_black_thunder, Naidu_bhstar, Rusakov_JWST_2025, degraaff_a_remarkable, Taylor_2025_capers}. In the BH$^*$ model (so named by \citet{Naidu_bhstar}), the UV excess is attributed to the host galaxy, which consists of a young stellar population with $M_* \lesssim 10^{9} M_\odot$. This could explain why variations in the UV have not been detected, as they would not be related to AGN activity. Additionally, dust is no longer needed to explain the red part of the SED, which is consistent with the non-detection in ALMA observations.  \\

Another question that remains open to this date is the evolution of LRDs. Selection criteria used in the literature predominantly identify objects in the range \( 4 < z < 8 \), regardless of whether the selection method is based on rest-frame properties or not \citep{kocevski_rise_2024, matthee_little_2024, barro_photometric_2024, akins_cosmos-web_2024, kokorev_census_2024, labbe_uncover_2023}. Because the redshift distribution exhibits a drop at \( z \approx 4 \) \citep{kocevski_rise_2024, Ma_counting_LRD_2025, Zhuang_nexus}, even in rest-frame selections such as the V-shape criterion used in \citet{kocevski_rise_2024} and \citet{euclid_LRDs}, at least one of the selection criteria must vary with redshift. Consequently, known LRDs must evolve as a function of redshift. This implies that LRDs evolve into a different state that no longer meets the selection criteria. If this assumption holds, there must be a transition phase in which these galaxies still exhibit LRD features while simultaneously developing new characteristics.\

Several scenarios can explain the evolutionary path of LRDs (see section \ref{sec:discussion}). However, recent observations suggest that some LRDs exhibit a resolved young stellar component, forming an extended structure beyond their initial compact core \citep{rinaldi_not_2024, chen_host_2024, Torralba_Lyalpha, Zhuang_nexus}. Since this stellar component may evolve over cosmic time, we investigate in this paper a specific evolutionary pathway in which LRDs acquire a young stellar component—observed as a blue star-forming periphery—while retaining their compact red inner core. To explore this scenario, we select a sample of 55 galaxies at a median redshift of \( z_{\text{med}} = 3.5 \), analyze their properties, and compare them to the known LRD population. We examine their potential evolution across redshift, aiming to shed light on the observed decline in the typical LRD selection around \( z \approx 4 \).

This paper is organized as follows: In Section \ref{sec:data}, we present the JWST/NIRCam  and HST/ACS data and the CEERS field, along with our sample selection from the catalog of \citet{merlin_catalogue}, intended to be related to LRDs at lower redshifts. In Section \ref{sec:inferring}, we present the methodology used to infer physical properties. The characteristics of the selected galaxies, including common features and differences with LRDs are presented in Section \ref{sec:relLRD} and Section \ref{sec:building}. We discuss different evolutionary scenarios for LRDs in Section \ref{sec:discussion} and conclude our study in Section \ref{sec:conclusion}. Throughout this work, we adopt an initial mass function (IMF) by \citet{chabrier} and the cosmological parameters from \citet{Planck_collaboration_cosmo_param}. Magnitude used are in the AB system \citep{OkeJB_abmag}.

\section{Data and Sample Selection}\label{sec:data}

\subsection{CEERS data}

In this work, we use JWST data from the Cosmic Evolution Early Release Science (CEERS) survey \citep{Finkelstein_CEERS, Finkelstein2025}, in the Extended Groth Strip (EGS) field. We specifically used the 10 JWST/NIRCam \citep{Rieke_Nircam_2003, Rieke_Nircam_2005, Beichman_NIRCAM_2012, Rieke_nircam_2023} pointings that covered approximately \( 97 \ \text{arcmin}^2 \) of the sky. Seven filters are available: F115W, F150W, F200W, F277W, F356W, F410M, and F444W with a depth at 5$\sigma$ for point sources of 29.15, 29.00, 29.17, 29.19, 29.17, 28.38, and 28.58 mag, respectively. We complete our short-wavelength coverage with the ACS/WFC instrument of the Hubble Space Telescope (HST) using the F606W and F814W filters \citep{Davis_HST_2007, Ford_1998_hubble}, including extensive data from CANDELS \citep{grogin_candels_2011,koekemoer_candels_2011}, with depths of 28.73 and 28.50 mag, respectively.

Concerning the observations and data reduction process, we refer the reader to \citet{Bagley_reduction_2023}. Briefly, the CEERS1, CEERS2, CEERS3, and CEERS6 pointings have been calibrated using the JWST Calibration Pipeline version 1.7.2 and CRDS pmap 0989. The remaining six pointings (CEERS4, CEERS5, CEERS7, CEERS8, CEERS9, and CEERS10) were calibrated using version 1.8.5 of the pipeline and CRDS pmap 1023. Then, all the pointings were reduced using stages 1 and 2 of the pipeline with additional steps such as removal of 1/$f$ noise, wisps, and distortion correction. The final mosaics were aligned through stage 3 using astrometry data from Gaia-EDR3 \citep{Gaia_astrometry_2O21}. Each pointing is background-subtracted and has a pixel scale of 0.03 $\text{arcsec} \ \text{pix}^{-1}$. All 10 pointings were then PSF-matched to the F444W-band image (0.16" FWHM), see \citet{carlos_dusty} for details. The total area of JWST/MIRI in the CEERS survey is $\approx 9$ arcmin$^2$ \citep{CEERS_miri1, CEERS_miri2}. Since we target a population with a surface density of $\approx 0.6-1 \ \text{arcmin}^2$ \citep{kocevski_rise_2024, perez-gonzalez_what_2024}, we choose to focus only on the JWST/NIRCam coverage to increase our statistics and have an homogeneous sample.

\subsection{CEERS catalog and pre-selection}
In this work, we use the catalog constructed by \citet{merlin_catalogue}, which includes data from multiple deep JWST extragalactic fields, including the CEERS field. Briefly summarizing the detection and flux measurement process, this photometric catalog was designed to maximize object detection, particularly for very faint galaxies across all JWST fields, using SExtractor v2.8.6 \citep{bertin_sextractor}. The detection image combines F356W and F444W bands, with an initial signal-to-noise (S/N) threshold of $\approx 2$. Photometry was extracted using Kron elliptical apertures \citep{kron_aperture} defined in the F356W+F444W detection image. The measured fluxes in all bands are then corrected using the aperture factor, determined from the detection image. A total of 82,547 objects were detected in the CEERS field, with available fluxes spanning the F606W to F444W filters and basic morphological parameters, including half-light radii (\( R_e \)), ellipticity, and position angle. Photometric redshifts were derived through spectral energy distribution (SED) fitting using \texttt{zphot} for one estimate \citep{Fontana_zphot}, and \texttt{EAzY} \citep{Eazy_z} across three independent runs. For further details, see \citet{merlin_catalogue}. Each source thus has four independent photometric redshift estimates.

From this catalog, we first apply some pre-filters. First, all bands must be available: F606W, F814W, F115W, F150W, F200W, F277W, F356W, F410M, and F444W (i.e., the error in each filter must be different from -99), which excludes sources at the very edges of the maps that are not measured in each filter. This removes $16,070$ galaxies out of $82,547$. A second threshold was applied to the signal-to-noise ratio, $S/N_{\text{F444W}} > 10$ and $S/N_{\text{F356W}} > 10$, to select well-detected objects (which retains $22,376$ sources). A final criterion is applied to select bright objects in the reddest band (F444W $< 27$ mag), targeting massive objects, similarly to LRDs \citep{barro_photometric_2024}. At this stage, our sample contains $18,218$ objects. Throughout this work, the redshift used for each object corresponds to the median of all photometric redshifts reported in the catalog of \citet{merlin_catalogue} when the spectroscopic redshift is not available.

\begin{figure}[h]
    \centering
\includegraphics[width=7cm,keepaspectratio]{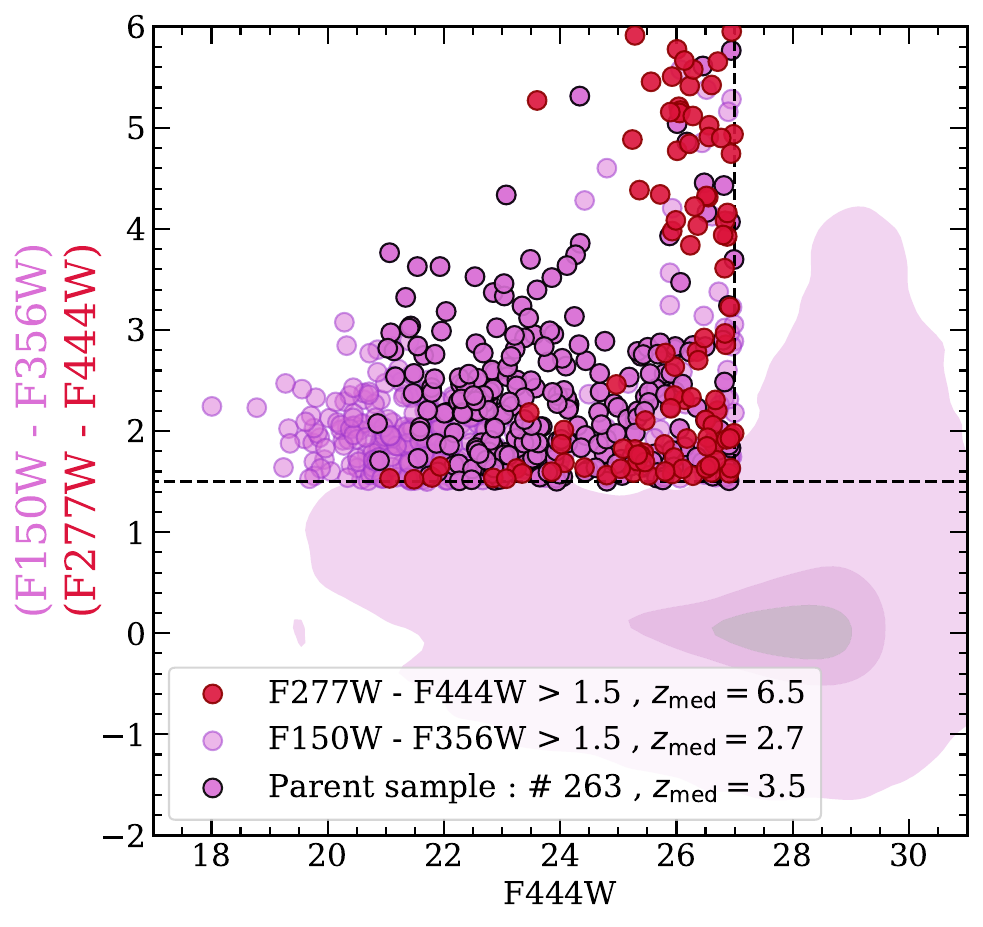}
\caption{ Color magnitude diagram illustrating the parent sample : in purple we show the sources that respect the color-magnitude criterion, and black edges denotes the parent sample, respecting $z>2.5$. In red, we show the color-magnitude selection for the red color of LRDs F277W - F444W > 1.5 used in \citet{akins_cosmos-web_2024}, having a $z_{median}\approx 6.5$. Red and purple contours represent all CEERS sources at 25\%, 50\% and 80\% levels shown for F150W-F356W.}
\label{fig:color_selection}
\end{figure} 

\begin{table*}[t]
\centering
\caption{Parameters used in \texttt{CIGALE}, the default value is used for all other parameters.}
\begin{tabular}{l l l l}          
\hline                       
    \hline
    Modules                                   &          Parameters                  &        Symbol      &         Values        \\ \hline \hline
    Star Formation History        &  e-folding time             &     $\tau_{\text{star}}$ [Myr]&           [1-100000], log step \\ 
    \texttt{sfhdelayed}, SFR $ \propto t \exp (-t/\tau)$  &  age\_main & $t_{\text{star}}$ [Myr]&                  [50-12000], step = 100\\ \hline
    Initial stellar population             & Initial mass function        &                   &    \citet{chabrier}              \\ 
    \texttt{bc03}     \citep{bruzualcharlot}    &  Metalicity                        &     $Z$           &           0.004,0.008,0.02,0.05\\ \hline
    Gas Emission                            &  Ionisation parameter             &    $\log U$         &           -2\\
    \texttt{nebular}                           &  Gas metallicity                 &   $Z_{\text{gas}}$       &           0.02\\ \hline
    Dust attenuation                &  reddening                       &  $E(B-V)$     &           [0-3.65] , step = 0.07\\ 
    \texttt{dustatt\_modified\_starburst}     &  ratio continuum/emission line & $E\_BV\_factor$  &  0.44                \\ 
                     \citep{calzetti}     &  &   &                  \\ 
\hline 
\end{tabular}  
\label{table:cigal_param}
\end{table*}

\subsection{Sample selection }\label{sec:sample_selection}

The number density of LRDs drops at $z<4$ \citep{kocevski_rise_2024, Inoyashi2025_LRDs,Ma_counting_LRD_2025, Zhuang_nexus, Pacucci_2025}, indicating that detections of these objects become rare at lower redshift. This implies that the selection criteria used to identify them may no longer apply at lower redshifts, suggesting that some of the known properties of these galaxies are likely to evolve over time. As these properties change, new characteristics may emerge that could hinder their detection using current methods. Here, we aim to identify galaxies that exhibit a compact red inner core alongside a blue, star-forming periphery. If such an evolutionary pathway is plausible for the LRD population—where LRDs acquire a blue, star-forming outskirts—there should exist a transitional phase in which the overall flux of the galaxy remains dominated by the characteristic red color of LRDs. There are several color selections for the red part of LRDs in the literature: F277W-F444W > 1.5 \citep{akins_cosmos-web_2024, perez-gonzalez_what_2024} and  F200W - F356W > 1.0 \citep{kokorev_census_2024, labbe_uncover_2023} to select LRDs with $z<6$. F200W-F444W is also used to probe the optical slope from $z=4.5$ to $z=7$ in \citet{barro_photometric_2024} and \citet{barro_extremely_2024}.

Therefore, the bands typically used to identify the red color in LRDs span from F200W to F444W. To select lower-redshift objects exhibiting similar rest-frame red colors, we adopt the analogous color criterion F150W - F356W > 1.5. Indeed, at \( 4 < z < 5 \), where the number density of LRDs has already started to decrease in most samples \citep{kocevski_rise_2024, barro_photometric_2024, akins_cosmos-web_2024, perez-gonzalez_what_2024,Ma_counting_LRD_2025}, F444W probes rest-frame \( 0.74-0.88 \ \mu\)m. To trace similar wavelengths at \( 3 < z < 4 \), we use F356W that covers \( 0.71-0.89\ \mu\)m versus F444W's \(0.89-1.1\ \mu\)m range. The same logic applies when substituting F200W with F150W: F200W probes $0.3-0.4 \ \mu$m rest-frame at $4<z<5$, and is substituted by F150W that covers 0.3 - 0.375 $\mu$m at $3<z<4$. Applying this first color selection selects $491$ galaxies with \( z_\text{med}= 2.7 \).

\begin{figure}[h]
    \centering
\includegraphics[width=8cm,keepaspectratio]{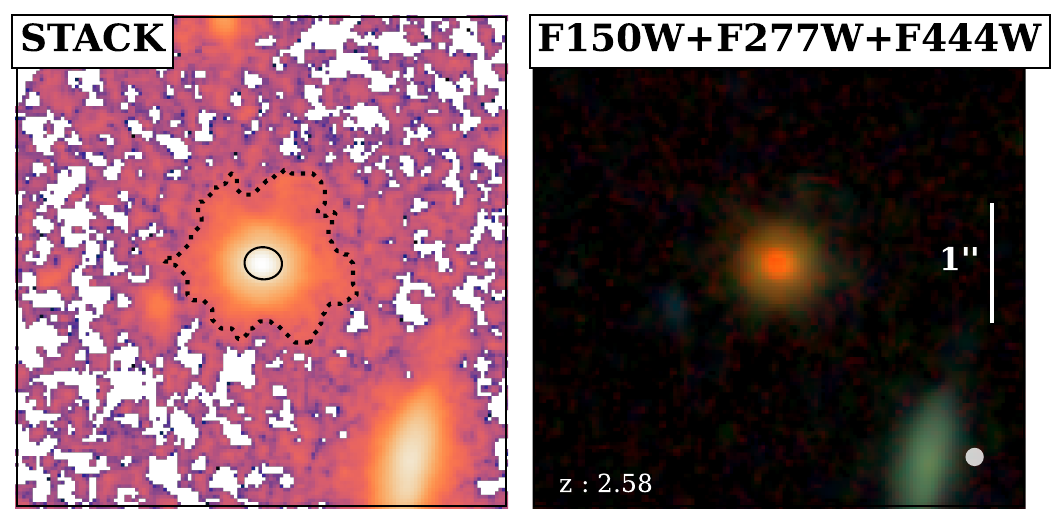}\caption{ Example illustrating the definition of the outskirts and inner regions.  
Left panel: stacked image combining all JWST bands, defined as STACK = F115W + F150W + F200W + F277W + F356W + F410M + F444W, using PSF-matched data. The dotted line shows the segmentation map, while the solid line traces the ellipse with a semi-major axis of $0.16^"=  2\text{HWHM}$.  
Right panel: color composite image using F150W (blue), F277W (green), and F444W (red). The white circle indicates the PSF FWHM beam at F444W. As expected, the galaxy exhibits a centrally concentrated red emission and a bluer, more extended periphery.
}
\label{fig:inner_outskier_11373_8}
\end{figure}

The redshift distribution of LRDs shows a rapid decline below $z\approx4$, indicative of significant evolutionary changes. To properly study this transition phase, we must focus on systems sufficiently close to $z\approx4$, as more distant (too low redshift) selections may yield galaxies that no longer share key characteristics with LRDs. A redshift cut is then applied, selecting $263$ objects with $z>2.5$ as shown in Figure \ref{fig:color_selection}, forming the parent sample. This criterion also serves to constrain the redshift range, preventing the filters from probing significantly different rest-frame wavelengths. Following visual inspection to remove artifacts and overly blended sources, 226 objects then remain with  \( z_{\text{med}} = 3.6 \). By removing blended objects, merger phases may be excluded from this study, as such systems often exhibit complex morphologies, as discussed in Section~\ref{sec:discussion}.

To probe the LRD population, a commonly adopted compactness criterion requires that the $4.4 \ \mu$m flux be predominantly concentrated toward the center. This criterion is slightly different among the literature and can be: \( F_\mathrm{F444W}(0.5'') / F_\mathrm{F444W}(0.2'') < 1.5 \) \citep{barro_photometric_2024}, \( F_\mathrm{F444W}(0.4'') / F_\mathrm{F444W}(0.2'') < 1.7 \)  \citep{ greene_uncover_2024, kokorev_census_2024, xiao_nocII_2025} or \( F_\mathrm{F444W}(0.2'') / F_\mathrm{F444W}(0.5'') > 0.5 \)  \citep{akins_cosmos-web_2024}, all leading to mostly unresolved objects. A population directly descendant from LRDs should similarly exhibit a compact core. To this end, we apply the condition \( F_{\mathrm{F444W,\ inner}} / F_{\mathrm{F444W,\ total}} > 0.4 \), where \( F_{\mathrm{X,\ inner}} \) denotes the flux measured within an elliptical aperture with a semi-major axis of \( 2 \times \mathrm{HWHM} = 0.16'' \) (half-width at half-maximum) centered on the object, and \( F_{\mathrm{X,\ total}} \) is the total flux measured within the segmentation map, in the filter X. To account for galaxy morphology, we compute the ellipticity as a function of radius using \texttt{Autoprof} \citep{Autoprof}, adopting the ellipticity measured at \( 0.16'' \). We adopt a major axis of $0.32''$, which is larger than the commonly used compactness threshold of $0.2''$, in order to apply a less stringent criterion that also allows partially resolved objects to be included. The segmentation map is generated from a detection image built by stacking all available JWST bands from F115W to F444W. This approach improves the signal-to-noise ratio and incorporates flux from extended, faint outer regions, enabling robust detection of low-surface-brightness structures (see Figure~\ref{fig:inner_outskier_11373_8}). This selection yields a sample of 110 objects. \\ 
Recent observations suggest that some LRDs exhibit a resolved young stellar component, forming an extended structure beyond their initial compact core \citep{chen_host_2024, Torralba_Lyalpha, Zhuang_nexus}. This young component may grow over cosmic time, potentially explaining why LRDs are no longer observed as a distinct population at lower redshifts. In this study, we explore an evolutionary pathway in which LRDs may develop a blue, star-forming periphery while retaining their red inner core. To probe such systems, two additional criteria are required to identify this star-forming outskirt. 

First, we impose \( \frac{F_{\mathrm{F115W,\ outskirt}}}{F_{\mathrm{F115W,\ total}}} > 0.4 \), ensuring that the rest-frame UV emission originates predominantly from the outer region, consistent with a younger stellar population relative to the core. Here, \( F_{\mathrm{X,\ outskirt}} \) is defined as the total flux minus the inner flux. A larger  threshold (above $0.4$) would fail to trace the phase in which LRDs begin to develop an outskirt, as the outskirts would already be prominent. Moreover, a too low threshold would probe objects without any outskirts and therefore pure LRDs that are not in this potential transition phase: $\approx 48$\% of the objects at a threshold lower than 0.4 are LRDs present in the \citet{kocevski_rise_2024} sample. This requirement reduces the sample to 72 objects.

Second, we apply a color cut of \( \mathrm{F115W}_\mathrm{outskirt} - \mathrm{F200W}_\mathrm{outskirt} < 2 \) to ensure the presence of young stars in the periphery. If such a criterion is not applied, we observe the presence of a strong Balmer break in the outer regions of some galaxies at \( z \approx 3.2 \). Such outskirts are unlikely to be associated with LRDs, and the Balmer break is therefore more likely to originate from an old stellar population, indicating a fully quiescent periphery. This contradicts the scenario of a recently acquired, star-forming envelope that we aim to probe in this work. Thus, applying this criterion limits the strength of the Balmer break in the redshift range \( z = 2\text{--}4.5 \), therefore constraining the presence of evolved stellar populations in the outskirts. Indeed, the Balmer break starts to affect the F200W filter at \( z \approx 4.5 \), meaning that a strong break can significantly boost the flux observed in F200W for galaxies at \( z \lesssim 4.5 \), making this band sensitive to older stellar components. In contrast, the F115W filter consistently probes rest-frame wavelengths shortward of the Balmer break down to \( z = 2 \), ensuring a reliable tracer of young stellar populations.

Applying only the first of the two criteria (\( \frac{F_{\mathrm{F115W,\ outskirt}}}{F_{\mathrm{F115W,\ total}}} > 0.4 \)) would allow fully quiescent systems with no UV emission in either region, while using only the second ( \( \mathrm{F115W}_\mathrm{outskirt} - \mathrm{F200W}_\mathrm{outskirt} < 2 \) ) would include galaxies whose cores are as blue as their outskirts, contradicting the scenario of a recently acquired star-forming envelope. 

After applying this selection, our final sample consists of 55 galaxies. A summary of the full selection process is provided below :
\[
\begin{aligned}& \hspace{2em}
\left\{
\hspace*{-0.8em} 
\begin{minipage}{0.9\textwidth} 
\begin{enumerate}[label={}, itemsep=8pt, parsep=0pt, topsep=0pt] 
    \item F150W - F356W > 1.5
    \item $z > 2.5$
    \item $\frac{F_{\text{F444W , inner}}}{ F_{\text{F444W, total}}}> 0.4$ \& $\frac{F_{\text{F115W , outskirt}}}{ F_{\text{F115W, total}}}> 0.4$
    \item $\text{F115W}_\text{outskirt} - \text{F200W}_\text{outskirt} < 2$
   
\end{enumerate}
\end{minipage}
\right.
\end{aligned}
\]
\vspace{0.25cm}

This selection relies on a single evolutionary scenario in which LRDs develop a blue star-forming outskirt while retaining their red inner core. As a result, the population probed in this work may not encompass all potential LRD descendants, and alternative evolutionary pathways may exist, as discussed later in Section \ref{sec:discussion}.
 
\begin{figure}[h]
    \centering
    \includegraphics[width=8cm,keepaspectratio]{ 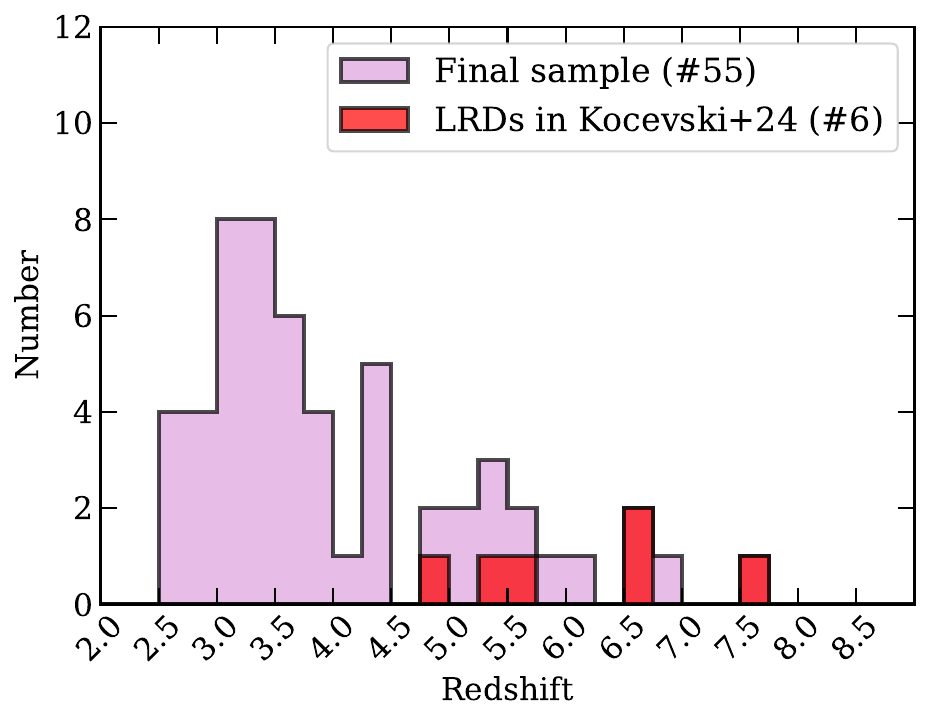}\caption{ Redshift distribution of the final sample in purple. 6 LRDs from our sample were also in \citet{kocevski_rise_2024} LRDs sample, shown in red.
}
    \label{fig:redshift_sample_distribution}
\end{figure}

\subsection{Final sample}\label{sec:final_sample}

Following the selection process, we retain a final sample of 55 objects, with \( z_{\text{med}} = 3.6 \pm 1.1\), each with a defined outskirt and inner part. Among these, 15 objects have spectroscopic redshifts. Six of them are already included in the LRDs sample studied in \citet{kocevski_rise_2024}, forming the high-redshift portion of our sample at \( z_{\text{med}} = 6 \) as visible in figure \ref{fig:redshift_sample_distribution}. All these LRDs present the same characteristic: an extended blue outskirt emitting in the UV, potentially indicating this transitional phase. Using SED fitting and morphological analysis, we aim to study the properties of the inner parts and the outskirts in the following sections. The goal is to identify features or characteristics within our sample that may be associated with LRDs at redshifts \( z < 4 \), in order to confirm or rule out a potential evolutionary link between the two populations.

Since our selection targets red galaxies, the sample may contain dusty star-forming galaxies (DSFGs), such as the optically-faint galaxies (OFGs), H-dropout or red star-forming galaxies \citep[RedSFGs,][]{Hdropout_wang_2019,carlos_dusty,gonzales_ceers_galaxy_evolution_2023,Tarrasse_2025} which typically exhibit a dust attenuation of \( A_\text{v} > 2 \). Our study focuses on objects with inner cores resembling those of LRDs, located around the transitional epoch at \( z = 4 \), and thus expected to be dust-free, consistent with the properties of LRDs \citep{labbe_uncover_2023, akins_cosmos-web_2024, casey_dust_2024}. SED fitting (see Section \ref{sec:inferring}) yields a median attenuation in the inner regions of \(\text{A}_{v,\text{med}} = 0.84 \pm 0.80\). Using the FIR catalog of Henry et al. (in prep), only three galaxies in our sample are detected in the \textit{Spitzer}/MIPS \(24 \, \mu\text{m}\) band with a S/N greater than 5, and none are detected in the \textit{Herschel}/PACS bands from \(100 \, \mu\text{m}\) to \(500 \, \mu\text{m}\) with an S/N above 3, indicating that our sources are not significantly dust-obscured. We also examined the \(H\)-dropout selection defined in \citet{Hdropout_wang_2019}, and found that only one source in our sample satisfies their criteria. This indicates that our sample has minimal overlap with this population.

\section{Inferring physical properties}\label{sec:inferring}

\subsection{Spectral Energy Distribution fitting}\label{sec:sed_fitting}

Like the LRD population, the inferred physical properties in our sample can highly depend on the model we assume. Indeed, the derived properties vary significantly depending on whether the modeling assumes a purely stellar component, a stellar component with an AGN, an AGN-only or the recently proposed BH$^*$ scenario \citep{perez-gonzalez_what_2024, akins_cosmos-web_2024, leung_exploring_2024, rinaldi_not_2024, degraaff_a_remarkable}. The nature of the AGN in LRDs remains an open question, here we adopt a stellar-only modeling approach. Consequently, all comparisons between the properties of our inner components and the characteristics of LRDs are based on this stellar-only assumption. We note that, the inferred mass of the inner region is potentially overestimated, as any AGN contribution is not accounted for.

To measure the flux of the inner part across all bands for constructing SED$_{\text{inner}}$, we measure the flux within the elliptical aperture defined by \texttt{Autoprof} in all filters (see Figure \ref{fig:inner_outskier_11373_8}). For SED$_{\text{outskirt}}$, we measure the remaining flux in the segmentation map mask after subtracting the inner elliptical aperture in all bands. Uncertainties were estimated by measuring the flux in 30 dark sky regions surrounding each galaxy, using the same aperture as that employed for the galaxy flux measurement. For this we used \(4.2^{\prime\prime}\)-wide cutouts centered on the target galaxy. The final uncertainty was taken as the root mean square (RMS) of these off-source measurements. \texttt{CIGALE} \citep{cigale} is then used to derive the physical properties of each region, employing identical models and parameters for both components: We assume a delayed star formation history (\texttt{sfhdelayed}) with SFR $\propto t \ \text{exp}(-t/\tau)$, a Chabrier IMF \citep{chabrier}, and \texttt{bc03} stellar populations \citep{bruzualcharlot}. We include both the \texttt{nebular} module and the \texttt{dustatt\_modified\_starburst} attenuation module, which combines the \citet{calzetti} law with extensions from \citet{leitherer} (see Table \ref{table:cigal_param}).

\begin{figure}[h]
    \centering
    \includegraphics[width=9cm,keepaspectratio]{ 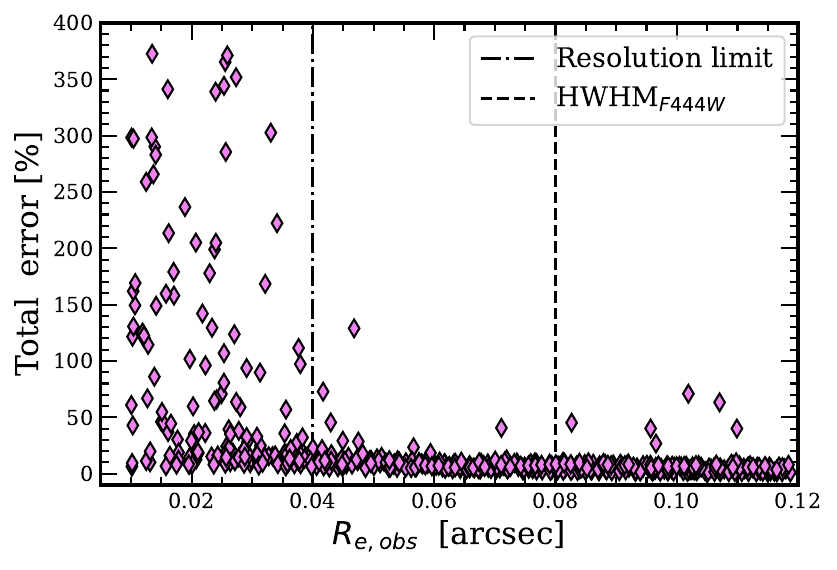}\caption{ Error in \% of the half-light radii $R_e$ and Sérsic index $n$ as a function of the inferred half-light radii for mock sources injected into dark areas of the CEERS field of view with mag = 26. The error is defined as $\text{Err}_\text{tot} =\sqrt{\text{Err}_{R_e}^2 + \text{Err}_n^2}$. The dotted line denotes the HWHM in F444W, whereas the dash-dotted line refers to the resolution limit adopted in this study.
}
    \label{fig:re_inferred_real}
\end{figure}

\subsection{Morphological analysis}\label{sec:morpho}

To infer morphological properties, we use \texttt{Astrophot} \citep{Stone2023} in single-band mode with the F444W filter of NIRCam. This software performs 2D profile fitting and takes into account the PSF, offering several fitting methods and models. We employ the \texttt{Levenberg-Marquardt} method with a single Sérsic profile for each galaxy. For this, we use small cutouts of 6 arcseconds, centered on each galaxy. All galaxies in the field of view were fitted to avoid contamination from neighbors. The Sérsic index $n$ was restricted to $0.2 < n < 10$, and the half-light radius to $0.01^{\prime\prime} < R_e < 1^{\prime\prime}$. Other parameters, such as the position $(x, y)$, the position angle $\theta$, the ellipticity $q$, and the central brightness density $I_e$, were left free. Among the sample, 30\% have $R_e < \text{HWHM}$, potentially leading to nonphysical characteristics.

To determine how far we can go below the PSF HWHM to study morphological features, we simulate 600 galaxies using single Sérsic profiles with \texttt{Astrophot}, convolving them with the PSF at F444W and injecting them into dark areas of the CEERS field of view. We fix the ellipticity at $q=0.7$, corresponding to the median values of our sample with $R_e > \text{HWHM}$. The injected half-light radius increases from  $R_e = 0.01^{\prime\prime}$ to \( R_e = 0.12 ^{\prime\prime}\) and the Sérsic index from $n=0.5$ to $n=6$. Dark regions are randomly selected in the field of view, and must have S/N<2. We also set the total magnitude of the injected sources to 26 mag, which corresponds to the faintest source among those with $R_e < \text{HWHM}$. Knowing the true values of \( R_e \) and \( n \) for the injected sources, we compute the total fitting error after retrieving their structural parameters using \texttt{Astrophot}:
\begin{equation}\label{eq:error}
    \text{Err}_\text{tot} =\sqrt{\text{Err}_{R_e}^2 + \text{Err}_n^2}=\sqrt{\left( \frac{R_{e,\text{obs}} - R_{e,\text{r}}}{R_{e,\text{obs}}} \right)^{2} + \left( \frac{n_\text{obs} - n_\text{r}}{n_\text{obs}} \right)^{2} }
\end{equation}

Here, \( R_{e,\text{obs}} \) and \( n_\text{obs} \) denote the inferred (observed) parameters obtained from fitting the mock sources, while \( R_{e,\text{r}} \) and \( n_\text{r} \) correspond to the true (real) input parameters used to generate the mock data. The results are shown in Figure~\ref{fig:re_inferred_real}. We find that below the HWHM, the total error remains low and nearly constant until a threshold of \( R_e \approx 0.04^{\prime\prime} \). Below this value, more than 40\% of the sources exhibit total errors exceeding 50\%, with a median error approximately four times higher than that of sources with \( R_e > 0.04^{\prime\prime} \). Adopting a lower resolution threshold would therefore risk mischaracterizing the morphology of the galaxies. However, this threshold can be considered a conservative resolution limit. Indeed, the most compact sources in our sample ($R_e <$HWHM) all have magnitudes brighter than 26 AB, and the effective resolution limit is expected to improve with decreasing magnitude \citep{akins_cosmos-web_2024}. Applying this resolution limit $R_\text{limit}=0.04$", we consider 77\% of our sample to be resolved in F444W.

\begin{figure}[h]
    \centering
    \includegraphics[width=9.5cm,keepaspectratio]{ 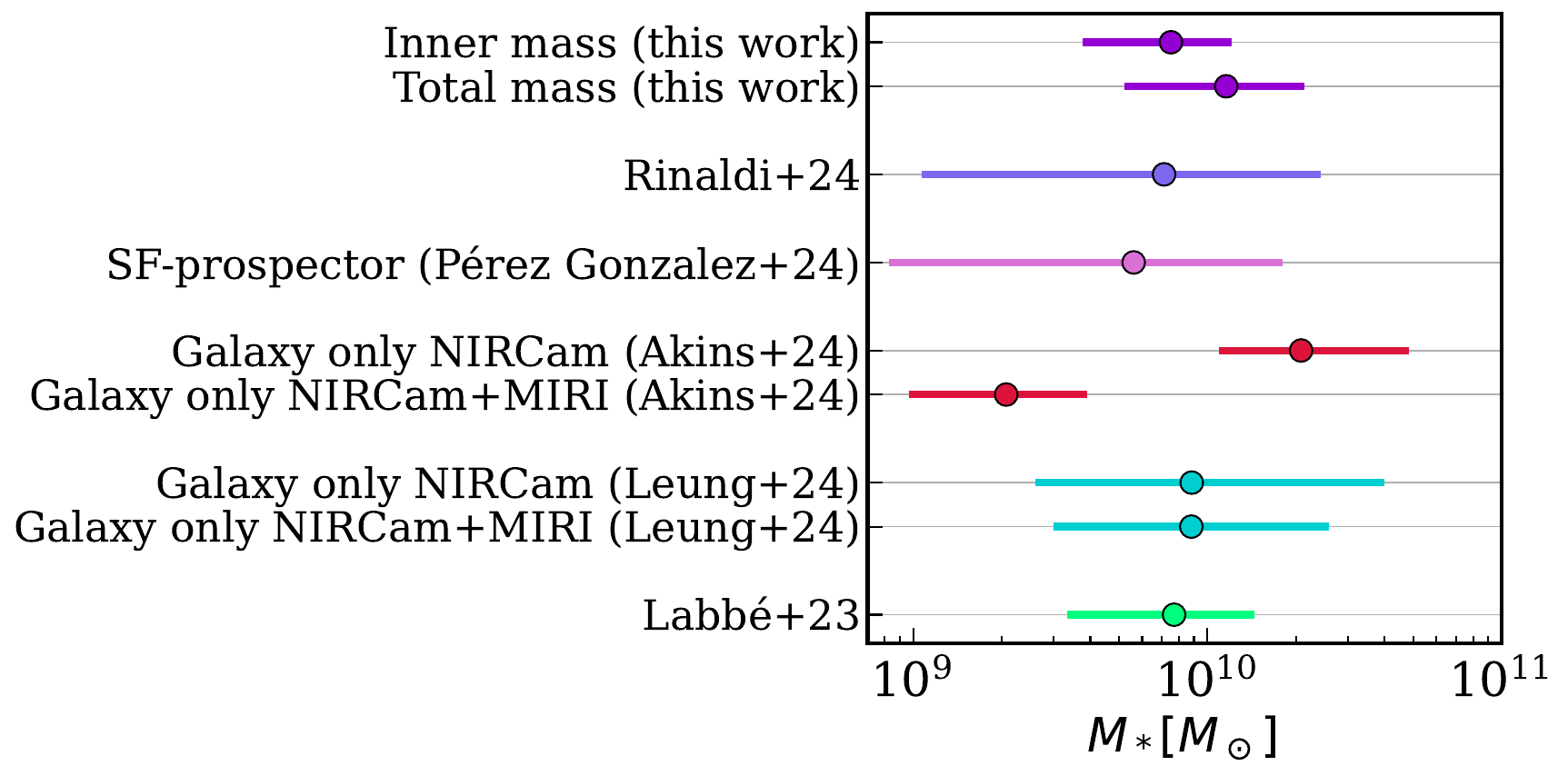}\caption{ Mass distribution for several samples of LRDs from previous works, assuming only a stellar model, from \citet{rinaldi_not_2024, perez-gonzalez_what_2024, akins_cosmos-web_2024, leung_exploring_2024, labbe_population_2023}. This sample is visible in purple for both the inner part and the total mass. }
    \label{fig:mass_distribution}
\end{figure}

\section{Relation with LRDs}\label{sec:relLRD}

\begin{figure*}[t!]
    \centering
    \includegraphics[width=14cm]{ 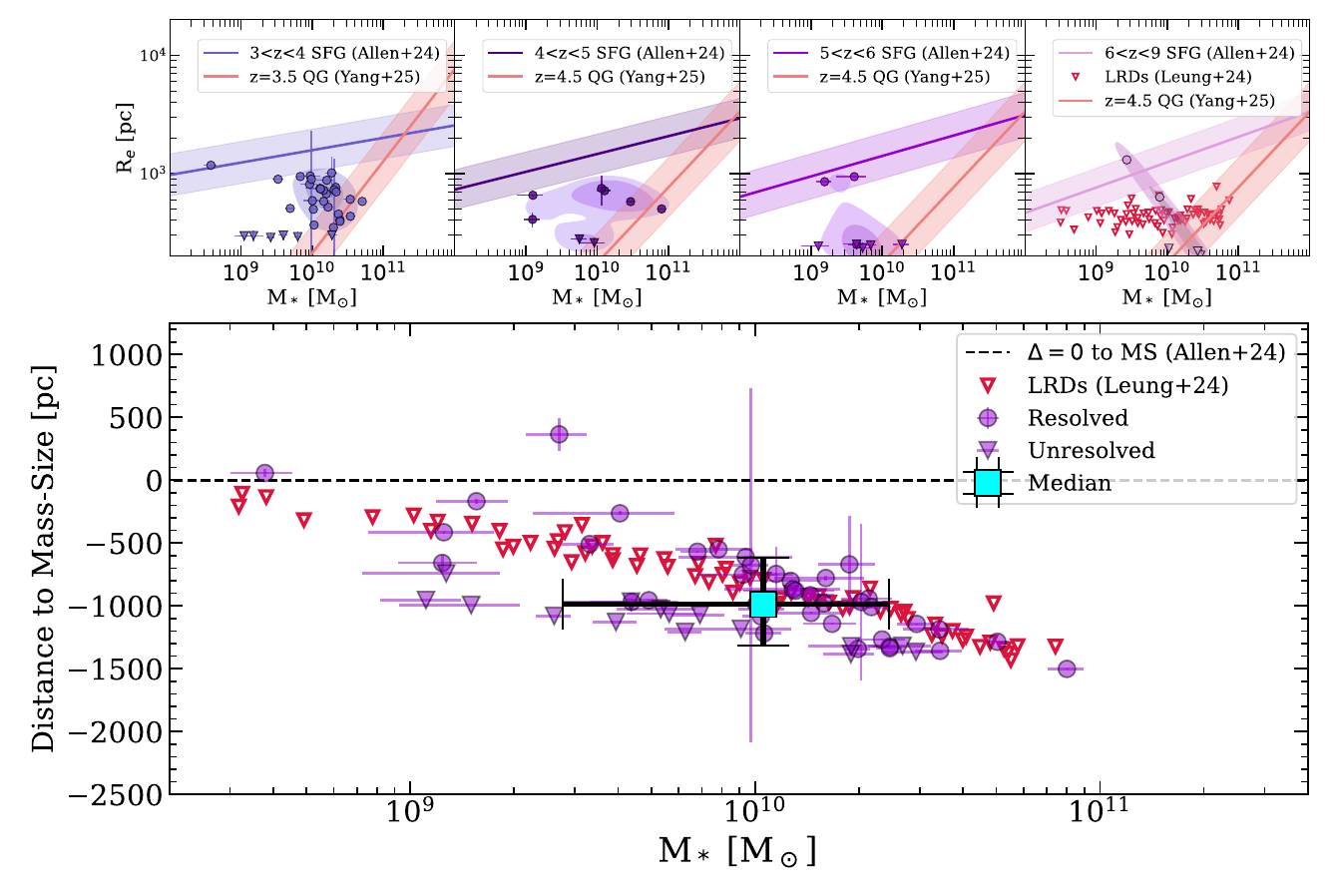}
    \caption{Top row: The stellar mass vs. \( R_e \) plane in each redshift bin of the sample, with the mass-size relation from \citet{allen2024} overlaid. The relation for quiescent galaxies (QG) from \citet{Yang_QG} is shown in orange. Triangular markers represent unresolved sources (for which we consider \( R_e \) = HWHM), while circular markers indicate resolved sources. Bottom: Distance to the mass-size relation for the entire sample. The cyan square represents the median of the sample, with 15\% and 85\% percentiles. The red triangles denote LRDs from \citet{leung_exploring_2024} which are also unresolved.}
    \label{fig:mass_size_relation}
\end{figure*}

This sample has been constructed to identify potential analogs of LRDs at lower redshifts, based on their color profiles and internal light distributions. To evaluate whether this sample is truly comparable to or related to LRDs, we first compare their stellar masses, although this comparison is subject to significant uncertainties, primarily due to the strong dependence of mass estimates on the physical model adopted in previous studies. Nevertheless, meaningful comparisons remain possible, provided that the same model is consistently applied. The masses of LRDs assuming only a stellar model are notably high, approximately  $10^{10} M_\odot $ as illustrated in Figure \ref{fig:mass_distribution}. This figure compiles several studies of LRDs that consider only the stellar model. The dark purple distribution represents the mass of the inner part and the total mass of the sample. The stellar masses of LRDs inferred without MIRI data by \citet{akins_cosmos-web_2024} are likely overestimated at \( z > 5 \), as shown by \citet{wang_miri_mass} and discussed in \citet{akins_cosmos-web_2024}. The inner stellar mass of our sample is consistent with the bulk of the distributions, with a median value of \( 10^{9.9\pm 0.4} \, M_\odot \), and a slightly higher total stellar mass of \( 10^{10.1 \pm 0.4} \, M_\odot \). This total includes the outskirt component of \( 10^{9.6\pm0.6} \, M_\odot \). The median stellar age of the outskirts is \( 300 \pm 200\, \text{Myr} \), indicating that they are effectively composed of young stars formed relatively recently as the selection has been constructed (See Section \ref{sec:sample_selection}). Given the median redshift of \( z_{\text{median}} = 3.6 \), this implies that the bulk of the outskirts began forming around \( z \approx 4.2 \).
 \\

Because galaxies are known to grow in size with redshift, following the mass-size relation \citep{vanderwel_2014, allen2024}, the compactness of this sample can be evaluated with reference to this relation in addition to our selection criterion ($\frac{F_{\text{F115W , outskirt}}}{ F_{\text{F115W, total}}}> 0.4$). As shown in Figure \ref{fig:mass_size_relation}, the mass-size relation is visible in each redshift bin described in \citet{allen2024} in the top row. The majority of the sample falls below the relation and can thus be considered compact. We then compute the distance to the mass-size relation in the bottom of Figure \ref{fig:mass_size_relation}. The median, shown in cyan, is  $990 \pm 330\ \text{pc}$ below the mass-size relation. Due to the unresolved sources in F444W (for which we consider $\text{R}_{\text{e}} = 0.04^"$), this can be regarded as a lower limit. In \citet{leung_exploring_2024}, the mass-size relation of LRDs was studied under both stellar-only and stellar+AGN model assumptions. Since we adopt a stellar-only model in our analysis, we show only the stellar results from their work in Figure~\ref{fig:mass_size_relation}, represented as red triangles. Since this sample was not resolved, the HWHM was used as $\text{R}_{\text{e}}$. The sample we constructed here exhibits a similar distance to the mass-size relation compared to LRDs, which is  $\approx 700 \ \text{pc} $. This indicates that this population of galaxies are as compact as LRDs relative to the mass-size relation. The median size of our overall sample is \( \text{R}_{\text{e,median}} = 520 \pm 300\ \text{pc} \) in F444W. If only the resolved sources are considered, it reaches \( \text{R}_{\text{e resolved,median}} = 670 \pm 230 \ \text{pc} \). For the unresolved sources, it is \( \text{R}_{\text{e unresolved,median}} = 255 \pm 30\ \text{pc} \), which is an upper limit.

\begin{figure}[h]
    \centering
    \includegraphics[width=9cm]{ 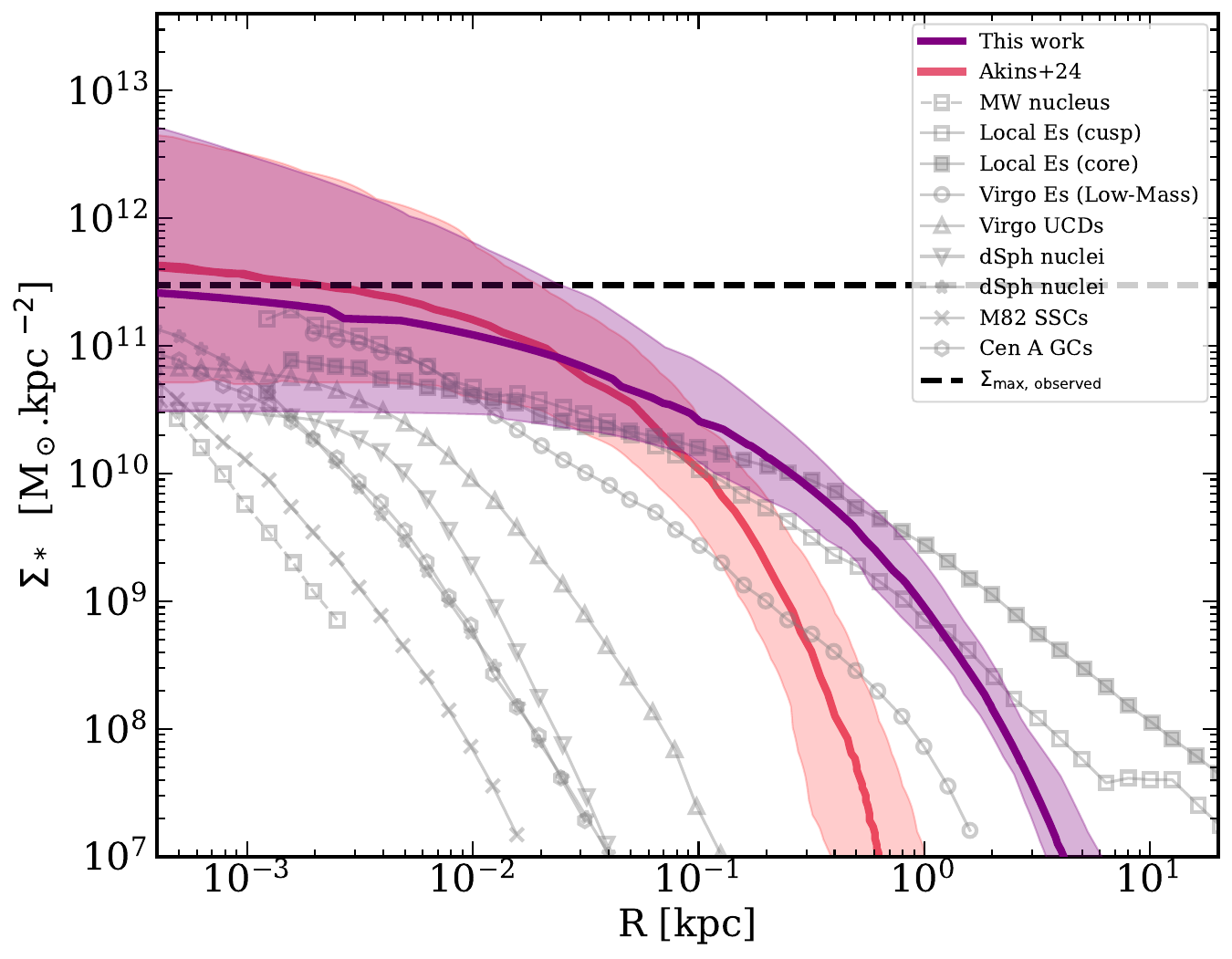}
    \caption{Stellar mass density profiles for our sample using only resolved galaxies in purple. In red the profile inferred by \citet{akins_cosmos-web_2024} assuming only a stellar model. We also add dense cores from \citet{hopkins2010} and \citet{grudi2019}.}
    \label{fig:sigma}
\end{figure}

Another feature of LRDs that has been studied is their stellar surface density \citep{akins_cosmos-web_2024, leung_exploring_2024, Guia2024, Khan_where_have_all_2025}. The central density inferred for LRDs is notably high, with values reaching $ \Sigma_* \approx 10^{11.6} \ \text{M}_\odot \ \text{kpc}^{-2}$ and $ \Sigma_* \approx 10^{11} \ \text{M}_\odot \ \text{kpc}^{-2}$ in \citet{akins_cosmos-web_2024} and \citet{leung_exploring_2024}, respectively. These values are broadly consistent with the maximum stellar surface density observed across a wide range of redshifts at $z=0-3$, as reported by \citet{hopkins2010} and \citet{grudi2019}, with $\Sigma_{\text{max}} = 3 \times 10^{11} \ \text{M}_\odot \ \text{kpc}^{-2}$. The sample studied here has a median redshift of $z_{\text{med}} = 3.6$. Thus, with the F444W band, we are directly tracing the $1\mu m$ rest-frame, effectively mapping the mass. Assuming that the surface brightness profile observed in the F444W band probes the surface mass profile, we use the previously inferred $R_e$ and $n$ from the single Sersic fit to derive the surface mass density profile. By knowing the total stellar mass $M_\text{tot}= M_\text{inner} + M_\text{outskirt}$, we can infer $\Sigma_e$, defined as the stellar mass surface density at $R_e$ : 
\begin{eqnarray}
      \Sigma_e & = &\frac{M_\text{tot}}{2\pi qR_e^2 J_n} \ , \label{eq:sigmae} \text{where} \\
      J_n & = & \int_{0}^{\infty} x \left( \exp \left[  -b_n x^{1/n} -1  \right] \right) \,dx\,
\end{eqnarray}

Where $q$ is the ellipticity, $J_n$ is then numerically computed. Because we are using morphological features and a non-negligible portion of our sample is unresolved in the F444W band, we compute this profile only for galaxies that are considered resolved, those with \( R_e > 0.04^{\prime\prime} \) (see Section \ref{sec:morpho}). Among the six LRDs in our sample, only one is resolved and therefore used to derive the surface density profile. To verify the robustness of the inferred $\Sigma_e$, we also calculated it directly using the mass-to-light ratio measured in the F444W band: $\Sigma_e = I_e \frac{M_*}{L_{F444W}}$, $I_e$ being the surface brightness at $R_e$, $M_*$ the stellar mass inferred with \texttt{CIGALE} and $L_{F444W}$ the total luminosity of the object in F444W.

\begin{figure}[h!]
    \centering
    \includegraphics[width=7cm]{ 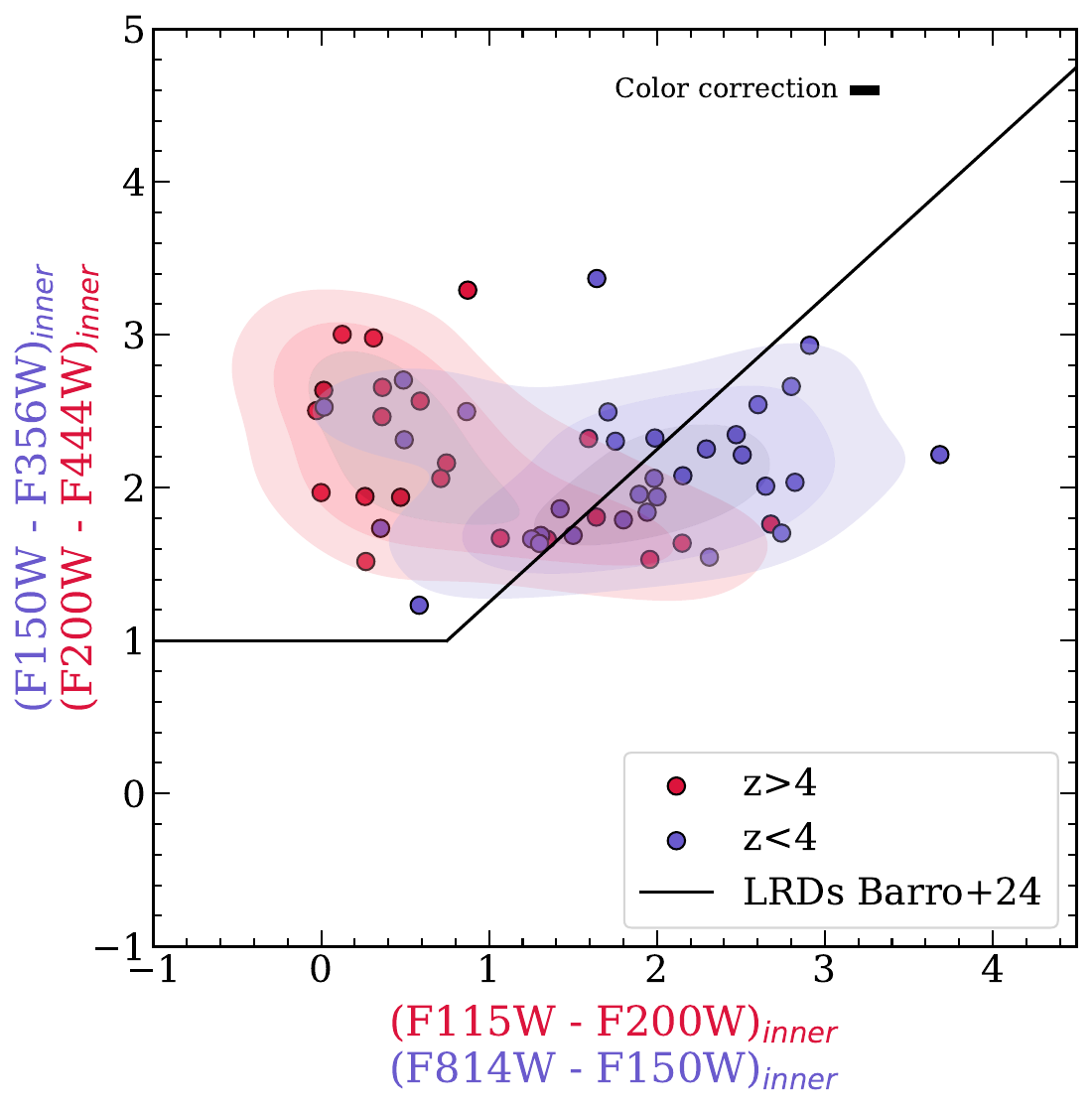}
    \caption{
Color–color selection of LRDs described in \citet{barro_photometric_2024}, originally defined for galaxies at \( z > 4 \), applied to the inner regions of our sample at \( z > 4 \) (shown in red). For galaxies at \( z < 4 \), the same rest-frame colors are probed by shifting the bands accordingly (shown in blue). Contours represent the 25\%, 50\%, and 75\% density levels for each distribution. The applied color correction is indicated by a thick black line, with an average shift of $\approx$ 0.2 mag.
}

    \label{fig:barro_selection}
\end{figure}

This yields consistent results: $ \Sigma_{*,\text{median}} = 10^{11.6\pm1.9} \ \text{M}_\odot \ \text{kpc}^{-2}$ in both cases. This method assumes a constant mass-to-light ratio across the galaxy, which is not the case in our sample. To quantify the bias of this assumption, we recompute the central mass profile using the inner part, which better describes the central region: $\Sigma_e = I_e \frac{M_{*,\text{inner}}}{L_{F444W,\text{inner}}}$. This yielded very similar results: $ \Sigma_{*,\text{median}} = 10^{11.5 \pm 1.9} \ \text{M}_\odot \ \text{kpc}^{-2}$.

In Figure \ref{fig:sigma}, we present the median profile of the stellar mass density in purple. The profile for LRDs, as reported by \citet{akins_cosmos-web_2024} and assuming only a stellar model, is shown in red. Additionally, the densest objects observed across a wide range of redshifts, as studied in \citet{hopkins2010} and \citet{grudi2019}, are also visible. The inferred stellar density profile aligns with the LRDs profile towards the center, suggesting that our sample has very similar physical properties in the inner region. More over, both profiles match the densest objects observed in the universe. However, the profile inferred for LRDs is slightly higher than $\Sigma_{\text{max}}$, which may be an overestimation in the case of LRDs due to their nearly unresolved size and the presence of an AGN, as discussed in \citet{akins_cosmos-web_2024}, which can also be a possibility in our sample. Furthermore, the physical size of our sample is larger than that of LRDs ($\approx 520 \pm 300$ pc vs 100-300 pc for LRDs), indicating a more extended structure which is also visible in Figure \ref{fig:sigma}.

\begin{figure}[h]
    \centering
    \includegraphics[width=8cm]{ 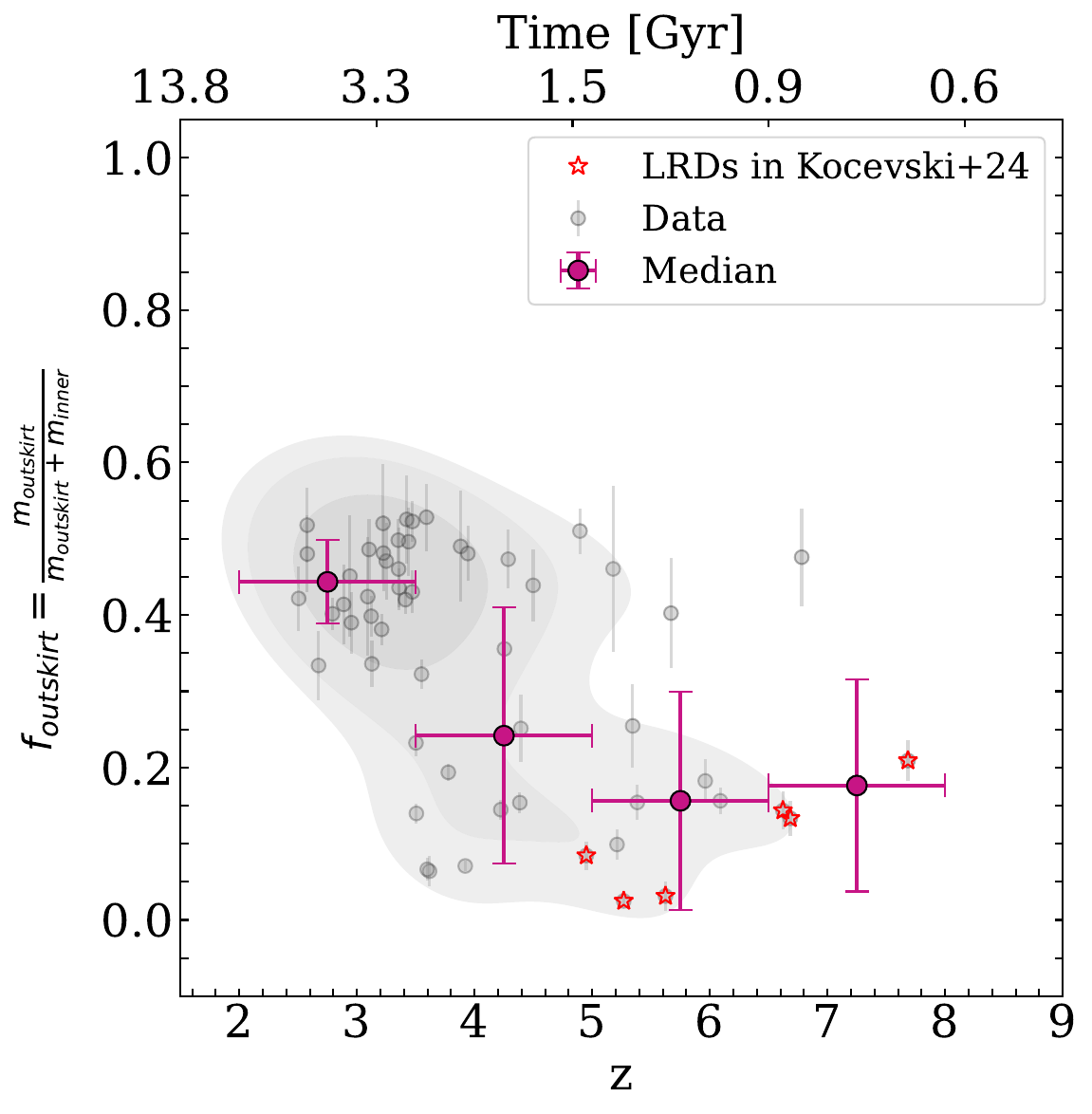}
    \caption{Outskirts mass fraction as a function of redshift. Gray dots represent the data, while gray contours illustrate their distribution at 0.25, 0.5 and 0.75 levels. The median is shown as a purple circle, with each point computed in bins of \( \Delta z = 1 \). Vertical error bars indicate the standard deviation, while horizontal bars represent the width within each bin. Red stars refer to the LRDs also in \citet{kocevski_rise_2024}.}
    \label{fig:outskirtgrowth}
\end{figure}

A selection based on a color-color diagram was developed by \citet{barro_photometric_2024} to identify both types of LRDs: those with and without the V-shape. However, at lower redshifts ($z < 4$), galaxies escape the selection due to the fixed bands used. To assess whether the inner cores of our objects are as red as LRDs while still satisfying the blue color criterion across a broad redshift range, we recompute the color–color diagram by shifting the filter bands for galaxies with \( z < 4 \). Figure~\ref{fig:barro_selection} illustrates this selection for the inner regions only, by shifting the color criteria as a function of redshift: in red for $z > 4$, using the same bands as \citet{barro_photometric_2024}: F115W-F200W and F200W-F444W. In purple, we show our sample with $z < 4$, with the bands shifted accordingly to trace the same rest-frame flux at $z<4$, i.e. using F814W-F150W and F150W-F356W. All the galaxies selected effectively present a red inner region, as our selection was designed to enforce. However, the blue color appears to evolve as redshift decreases. Specifically, the blue color at \( z < 4 \) (F814W-F150W) appears redder than its counterpart at \( z > 4 \) (F115W-F200W), causing approximately 50\% of the galaxies at \( z < 4 \) to fall outside the LRD color selection defined by \citet{barro_photometric_2024}. This shift suggests a possible cosmic evolution of the population over time.

\section{The building of an outskirt}\label{sec:building}

Our sample appears to be more extended than typical LRDs (see Figure~\ref{fig:sigma}), yet it retains a similarly dense and red core, and becomes slightly more massive when the outskirts are taken into account, suggesting a potential growth scenario. If the outskirts formed around LRDs in a later stage, we should observe a time correlation with the outskirt mass fraction. To investigate this, we compute the outskirt mass fraction, $f_{\text{outskirts}} = m_{\text{outskirts}} / (m_{\text{outskirt}}+m_{\text{inner}})$, as a function of redshift, as shown in Figure \ref{fig:outskirtgrowth}. There is a strong correlation with redshift, with the outskirt mass fraction increasing at lower redshifts. The LRDs from \citet{kocevski_rise_2024} present in our sample (visible as red stars) lie at the very low outskirt mass and high redshift end of the figure. This is in line with the idea of a growth scenario. The observed correlation could be affected by cosmological surface brightness dimming. Since dimming causes the surface brightness to appear fainter at higher redshift ($I_\text{obs} \propto I_0(1+z)^{-4}$), we can only detect the brightest outskirts as redshift increases (see Appendix~\ref{appendix_1}). In other words, at higher redshift we preferentially detect the most massive outskirts, potentially missing the low-mass ones, if any. Consequently, the outskirt mass fraction at high redshift shown in Figure~\ref{fig:outskirtgrowth} should be considered as upper limits. Once corrected for this effect, the trend would likely appear even clearer.

Because of this growth, the compactness criterion may no longer select these galaxies, also correlated with an evolution of colors at lower redshift (see Figure \ref{fig:barro_selection}). 

\begin{figure}[h!]
    \centering
    \includegraphics[width=7.5cm]{ 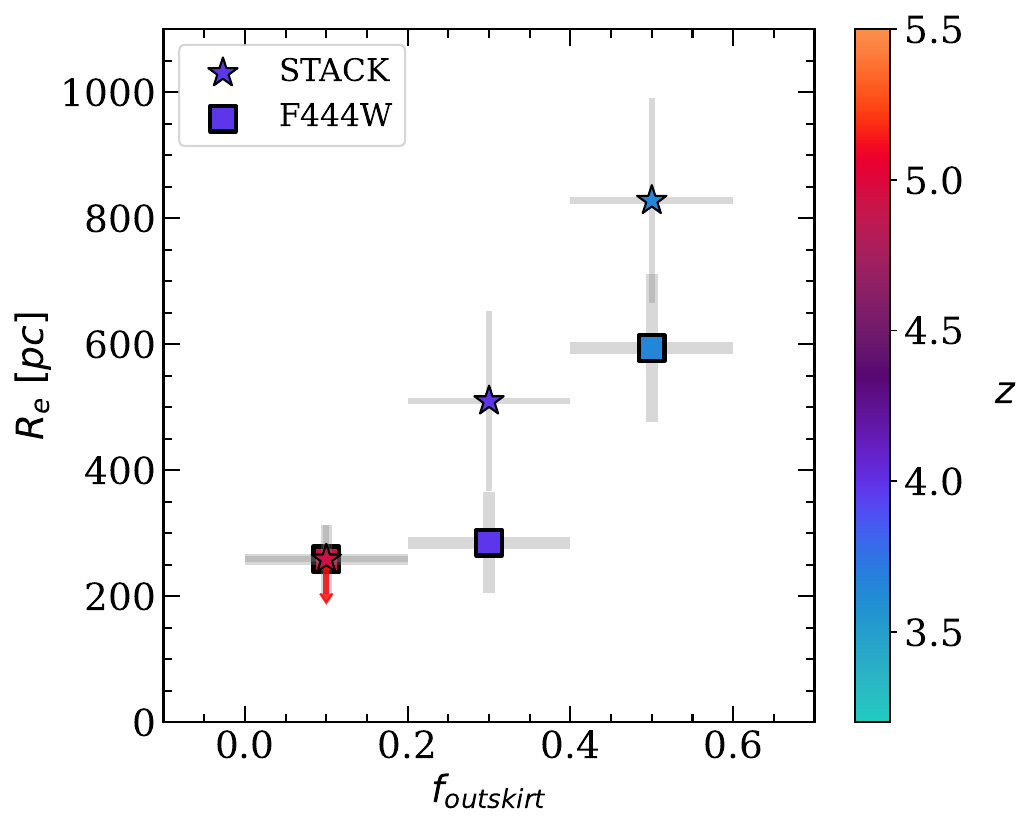}
    \caption{Half light radii $R_e$ as a function of $f_{outskirt}$. Stars correspond to the median \( R_e \) from stacking all the filters, and squares correspond to only F444W. The color bar shows the mean redshift in each \( f_{outskirt} \) bin. }
    \label{fig:re_foutskirt}
\end{figure}

To further investigate the spatial growth of the outskirts, we compute the median stacked image of our sample within bins of  $f_{\text{outskirt}} = [0-0.2, 0.2-0.4, 0.4-0.6] $ using the F444W filter. This stacking approach increases the $S/N$ by at least a factor of three in each \( f_{\text{outskirt}} \) bin, enabling the detection of faint substructures that would otherwise remain undetectable in individual images. In each bin, we fit a single Sérsic profile using \texttt{Astrophot} to derive \( R_e \) as a function of \( f_{\text{outskirt}} \), shown as squares in Figure~\ref{fig:re_foutskirt}. Since $R_e$ is not significantly affected by cosmological dimming \citep{Yu_desi_morphology, Ormerod_dimming}, and given that we increased our $S/N$ to minimize its potential impact, we consider our derived $R_e$ to be robust and physically meaningful.

As expected, as the mass fraction in the outskirts increases, the physical size of the galaxies also grows with decreasing redshift. The most compact systems, with the lowest \( f_{\text{outskirt}} \), are unresolved in F444W and are located at high redshift, with an upper size limit of  $260\pm55$ pc. This increases to around $285\pm80$ pc for galaxies with moderate \( f_{\text{outskirt}} \), and to about $590\pm 120$ pc for the most extended and massive systems. Stars in Figure~\ref{fig:re_foutskirt} represent the median sizes derived across all JWST PSF-matched bands, probing not only rest-frame NIR emission but also the rest-frame UV, allowing to increase the $S/N$ by a factor 9, allowing faint detection. Galaxies without any outskirt are still not resolved, having an upper limit of 260 pc, which is expected, and those with a moderate outskirt reached $510 \pm 140$pc while the most extended systems with the largest outskirt reached $828\pm 160$ pc. This suggests that the UV emission starts to be more prominent and detected when the fraction of outskirt is above 0.2, at redshift $\approx 4$.  

\begin{figure}[h]
    \centering
    \includegraphics[width=9.5 cm]{ 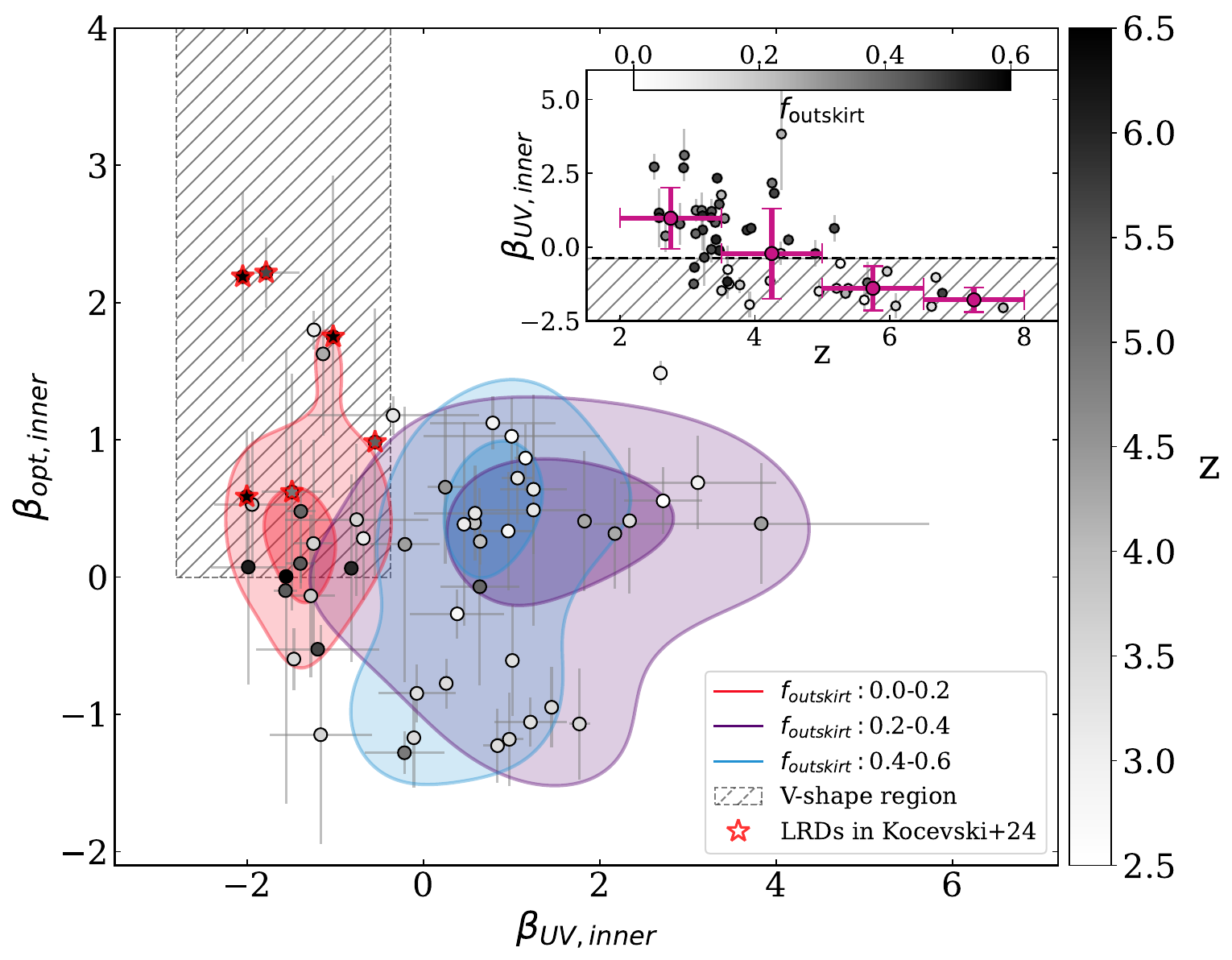}
    \caption{$\beta_{\text{UV}} / \beta_{\text{opt}}$ plane of the inner regions of the sample. The fraction of the mass outskirt is represented by contours with 0.4 and 0.8 levels, increasing from red to purple and blue. Redshift is indicated by a gray color bar. The LRD selection from \citet{kocevski_rise_2024} is shown in the light gray dashed region and LRDs in common are shown in red stars. In the top right-hand corner, we show $\beta_{\text{UV,inner}}$ as a function of redshift, the colorbar denotes the mass fraction of the outskirt $f_\text{outskirt}$, the shaded region refers to the V-shape selection.}
    \label{fig:buvbopt}
\end{figure}

A common way to select LRDs is by their V-shaped SED, based on continuum slopes, first used by \citet{kocevski_rise_2024}. To determine if the formation of these outskirts impacts the V-shape, we reconstruct the $\beta_{\text{UV}}$ vs. $\beta_{\text{optical}}$ plane, a rest-frame criterion, using only the inner SED which is expected to be mostly related to LRDs. We compute $\beta_{\text{UV,inner}}$ and $\beta_{\text{optical,inner}}$ assuming $f_{\lambda}\propto \lambda^{\beta}$ in the rest-frame UV and rest-frame optical region. Each slope $\beta_{\text{inner}}$ is compute using 3 bands blueward or redward of the Balmer break at $3645$\r{A} as in \citet{kocevski_rise_2024}, by adjusting a linear fit in log($f_{\lambda}$) space. Due to the shallower depth of the HST filters (28.73 and 28.50) compared to F115W (29.15), some of our sources are undetected in either F606W or F814W, occasionally resulting in negative flux values within the central aperture due to background subtraction. In such cases (affecting 5 objects), we compute \(\beta_\text{UV,inner}\) using only the two available bands and conservatively set the associated uncertainty to 1.

In Figure \ref{fig:buvbopt}, we display this plane and overlay the V-shape selection from \citet{kocevski_rise_2024} with a dashed region. Contours are also visible for different bins of $f_{\text{outskirts}}$. The red stars indicate the 6 LRDs also present in \citet{kocevski_rise_2024}. It appears that a low $f_{\text{outskirts}}$ is necessary to maintain the V-shape. As $f_{\text{outskirts}}$ increases with decreasing redshift, galaxies depart from this V-shape selection. In the top right-corner we display the evolution of $\beta_\text{UV,inner}$ as a function of the redshift: as the redshift decreases, $\beta_\text{UV,inner}$ increases and escapes the V-shape selection (dashed region), at $z\approx 4$. This departure could explain why even a rest-frame selection suffers from a drop around $z = 4$, due to physical evolution: as the outskirt is growing, a stellar component becomes non-negligible around this redshift for the majority of LRDs. If this stellar component erases the V-shape, it could also explain why some LRDs exhibit the V-shape while others do not. Broad color-color selection traces all LRDs, with or without a stellar component, and the V-shape may select only those dominated by AGN, consistent with the finding that 70\% of LRDs selected using the V-shape present broad lines, indicative of AGN \citep{kocevski_rise_2024}. \\

\begin{figure}[h]
    \centering
    \includegraphics[width=9cm]{ 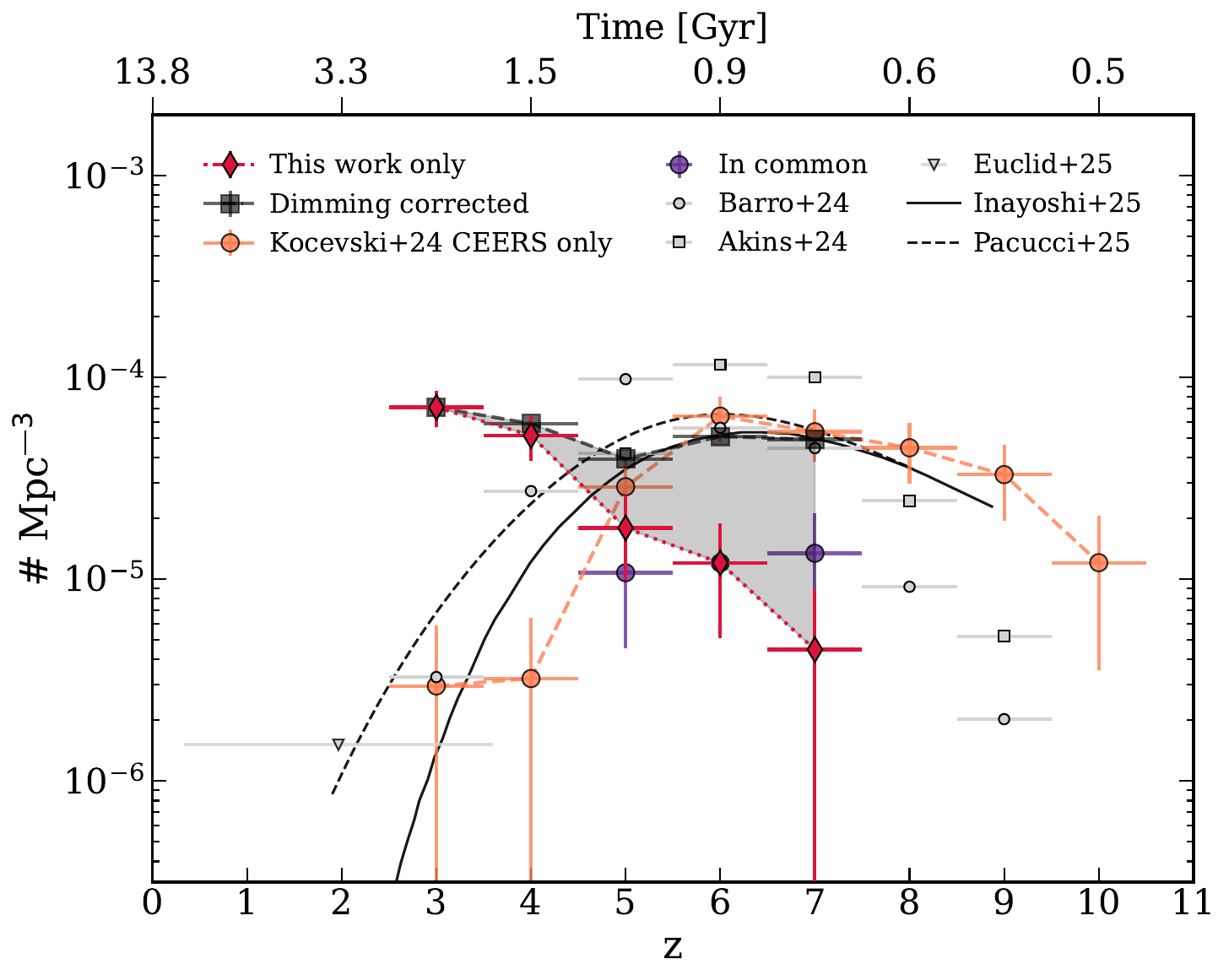}
    \caption{ Number density as a function of redshift from our sample visible in red, excluding the 6 common sources with the LRDs in CEERS from \citet{kocevski_rise_2024}, here shown in orange. In purple, the number density of the sources in both samples is shown. The dimming-corrected number density of our sample is shown with dark gray square symbols, interpreted as an upper limit. LRDs from \citet{akins_cosmos-web_2024}, \citet{barro_photometric_2024}, and \citet{euclid_LRDs} are represented by light gray squares, circles, and triangles, respectively. Horizontal error bars indicate the redshift bins used to compute the number density, and vertical error bars represent Poissonian errors. The theoretical evolution from \citet{Pacucci_2025} is shown as the dashed black line, while the solid line refers to the model of \citet{Inoyashi2025_LRDs}.
 }
    \label{fig:numer_density}
\end{figure}

If our evolution and growing scenario of LRDs is corect, the number density of our sample should be compatible with the one of LRDs and at lower redshifts. We compute the number density as a function of redshift in Figure~\ref{fig:numer_density}, by dividing our sample into redshift bins and calculating the corresponding comoving volume. As we have 6 LRDs in our sample, we exclude them to compare our distribution with the one of LRDs. Our sample is visible in red and the common sources with \citet{kocevski_rise_2024} in purple. We also compute the distribution of LRDs based on the sample of \citet{kocevski_rise_2024} in CEERS only (visible in orange), and the ones from \citet{akins_cosmos-web_2024}, \citet{barro_photometric_2024}, and \citet{euclid_LRDs}. Additionally, the two theoretical models of the number density developed by \citet{Pacucci_2025} and \citet{Inoyashi2025_LRDs} are shown for comparison. 

Firstly, we find that the final number density in the lowest redshift bin, \(  z=3\pm 0.5\), is \( 10^{-4.15\pm 0.09} \ \text{Mpc}^{-3} \), which is consistent with the number density of LRDs at \( z = 6\pm0.5 \) in CEERS (based on the sample of \citet{kocevski_rise_2024}),  \(  10^{-4.17 \pm 0.11} \ \text{Mpc}^{-3} \). The fact that we recover a number density consistent with that of LRDs suggests that this evolutionary scenario may be viable and potentially the most likely pathway for the evolution of LRDs. Nevertheless, this does not exclude other possibilities, as some LRDs in a merger state have already been observed \citep{Merida_LRDs_merger, Tanaka_LRDs_merger}, as discussed in Section \ref{sec:discussion}.

Secondly, the number density of LRDs (in orange) declines at the same redshift where our distribution (in red) starts to increase, around \( z \approx 5 \). However, since we select objects with a recently formed blue periphery, this feature may either be undetectable at higher redshifts—due to cosmological surface brightness dimming—or may not have formed yet. To account for this effect, we applied a correction to the observed number density in each redshift bin, corresponding to the fraction of galaxies likely missed due to cosmological dimming ( see Appendix \ref{sec:appendix}). The corrected number density is shown as dark gray square symbols in Figure~\ref{fig:numer_density}. We emphasize that this correction assumes all our sources would already possess a developed outskirts component at higher redshift. In this case, the apparent emergence of the outskirts at lower redshift would not be a physical effect, but rather an observational bias caused by cosmological dimming—implying that LRDs could host undetected peripheries at high redshift. However, stacked analyses by \citet{akins_cosmos-web_2024} and \citet{Delvecchio_stack}, who respectively combined 429 and 302 individual LRDs from different samples across filters ranging from F150W to F444W, did not detect any extended emission in their stacked images. This suggests that the majority of LRDs are intrinsically compact and do not host faint, dimmed outskirts that would remain undetectable in individual observations. Therefore, this favors the physical formation of a stellar periphery over cosmic time rather than an observational bias caused by cosmological dimming. Based on this, we conclude that the dimming-corrected number density should be considered as a conservative upper limit, while the true number density of such objects is likely closer to the observed values. It is also important to mention that it does not exclude other evolutionary scenario, as some LRDs in a merger state have already been observed \citep{Merida_LRDs_merger, Tanaka_LRDs_merger}, as discussed in Section \ref{sec:discussion}, it rather suggest that the other evolutionary scenario may not be the main channel of evolution for LRDs.

V-shaped LRDs might be a pure black hole embedded in dense gas as recently suggested by \citet{Ji_black_thunder, Inoyashi2025_LRDs, Naidu_bhstar, Rusakov_JWST_2025}. As the V-shaped SED diminishes with the growth of the stellar component and time (See Figure \ref{fig:outskirtgrowth}), the influence of this component likely increases, and LRDs begin to lose their known characteristics. In this scenario, V-shaped LRDs represent an earlier phase, while LRDs without the V-shape represent a later one. \\

\section{Discussion}\label{sec:discussion}
\begin{figure}[h]
    \hspace{-0.4cm}
    \includegraphics[width=9cm]{ 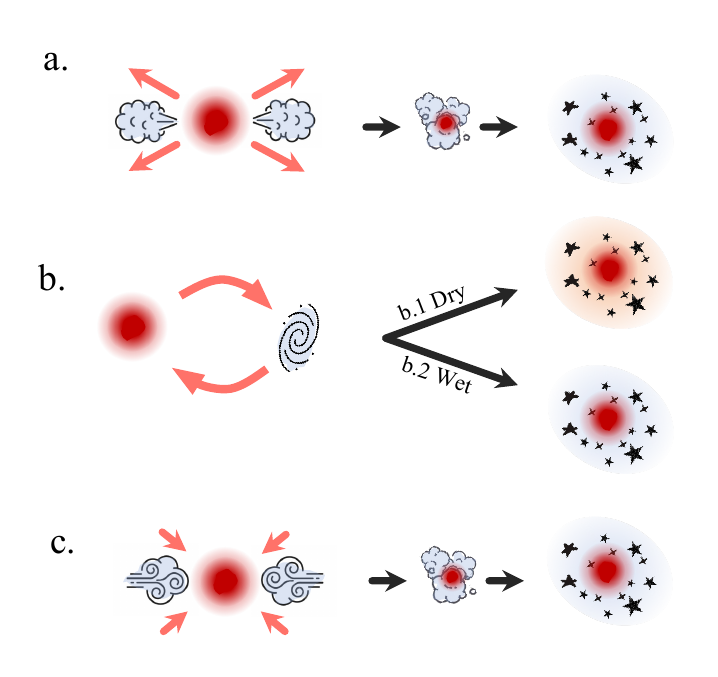}
    \caption{Drawing illustrating the different possible scenarios of evolution of LRDs, as discussed in Section \ref{sec:discussion}: through (a) feedback, (b) mergers, or (c) accretion. }
    \label{fig:doodle}
\end{figure}

We now propose to discuss the origin of the growing stellar outskirt.
Assuming that these galaxies are initially LRDs, several scenarios could take place, allowing stars to populate the outer regions, as illustrated in Figure \ref{fig:doodle} :
\begin{enumerate}[label=\alph*)]
    \item Gas can be expelled from AGN/supernova feedback from the inner region and then form stars.
    
    \item One or a succession of mergers can favor the formation of the outer region:
    \begin{enumerate}[label=(b).{\arabic*},ref=(b).{\arabic*}]
        \item Wet mergers can bring gas that fuels the LRDs through cosmic time.
        \item Dry mergers, however, are more likely to form an outer region solely through kinematic processes and tidal interactions, due to the absence of gas.
    \end{enumerate}
    
    \item Direct accretion from the environment by either cold streams or diffuse hot accretion may help to form the stellar outskirt.
\end{enumerate}

Throughout \textbf{scenario (a)}, the gas initially present in the inner regions could be ejected by feedback from the central AGN or supernovae, potentially forming a gaseous outskirt that may later give rise to new stars. This would result in younger metal-rich stars with higher sSFR in the outskirt and, consequently, significant opacity. In general, as discussed in \citet{Baggen_the_small_2024}, an expansion of the stellar distribution can occur with mass loss due to AGN feedback or stellar winds, increasing the size of the galaxy \citep{Fan_The_dramatic} by the redistribution of the gas in the galaxy \citep{Khan_agn_feedback_2025} and therefore could be applied to LRDs. The molecular gas reservoir of an LRD was measured by \citet{akins_tentative_2025} using CO(7-6) and $[C_\text{I}](2-1)$, where the amount of neutral molecular gas appeared to be limited $M_\text{neutral\ gas}<10^{10}M_\odot$. However, the presence of an extreme gas reservoir was suggested by \citet{Ji_black_thunder, degraaff_a_remarkable, Rusakov_JWST_2025} and as in the Black Hole Star model (BH$^*$) proposed in \citet{Naidu_bhstar}, in which the gas is extremely dense ($n_{\text{H}} \approx 10^{11} \ \text{cm}^{-3}$) and turbulent ($v_\text{turb}=500$ km$\text{s}^{-1}$). Due to these extreme conditions, the gas could be spatially redistributed because of instabilities favored by either AGN or supernovae feedback from the host galaxy, a scenario also discussed in \citet{Taylor_2025_capers}. However, the fraction of ionized gas $\chi = \frac{n_{\text{H}^+}}{ n_{\text{H}}}$ remains to be studied in such systems. A large fraction of ionized gas would also require an explanation of how this energetic component becomes neutral before forming new stars. Moreover, if the velocity of the gas expelled is too important ($v_\text{gas}> v_\text{escape}$), it could be expelled beyond the gravitational potential of the system, thus no longer remaining bound to it and preventing star formation in the outskirts. Therefore, although plausible, this scenario remains highly speculative.

\textbf{Scenario (b)} could also occur, with mergers between LRDs or between a LRD and a galaxy. If the merger involves a gas-rich component (scenario (b).1), it may bring gas in the outskirt through tidal events and kinematic interactions, leading to the formation of new stars outside the core. However, mergers generally tend to concentrate the gas in the central region \citep{Hopkins_inflows_merger} and therefore also trigger star formation in the inner part, so that a higher sSFR and younger stars in the center is more likely to be observed. The stars already in place in the LRD may also be ejected into the outskirt through dry merger events (scenario (b).2), as discussed in \citet{Khan_where_have_all_2025}, who modeled the evolution of LRDs over hundreds of millions of years, accounting for the impact of supermassive black hole (SMBH) mergers. Using initial parameters from LRDs first observed by \citet{Baggen_the_small_2024} at $z_{\text{spec}} = 6.68$ and $z_{\text{spec}} = 6.98$, they found that the galaxy undergoes mergers that flatten the initial stellar mass density profiles by one order of magnitude, decreasing from $10^{12} \ \text{M}_\odot \ \text{kpc}^{-2}$ to $10^{11} \ \text{M}_\odot \ \text{kpc}^{-2}$ over a timescale of $100-800$ Myr, which approximately corresponds to the period between $z = 4.5$ and $z = 3$. In the BH$^*$ model of \citet{Naidu_bhstar}, the blue component of LRDs is explained by a very young stellar population that would be approximately 1 Gyr old at $z \approx 3$ if formed at $z \approx 5$. If these stars are then redistributed through dry merger events, the result would be an outskirt as old as the inner region with similar colors. LRDs in a merger state have already been observed by \citet{rinaldi_not_2024} and \citet{chen_host_2024}, using high-resolution UV data, and a more complex system involving two LRDs merging in \citet{Merida_LRDs_merger} and \citet{Tanaka_LRDs_merger}, supporting this scenario (b) for at least some descendants of the LRD population.

In \textbf{scenario (c)}, the gas is directly accreted by cold gas filaments or diffuse accretion of warm gas (the so called hot accretion mode). Such accretion is expected to feed a disk together with its angular momentum \citep{Kere_mass_simulation_cold_inflows, Dekel_cold_streams, Dekel_2009}. Because LRDs are high redshift sources, the cold accretion mode is more likely to dominate those systems. The infalling gas would be metal-poor, leading to young stars with low metallicity in the outer region \citep{sanchez_gas_accretion, Pichon_2011}. This has also been discussed in \citet{kocevski_rise_2024}, where infalling gas is suggested to bring angular momentum, promoting star formation on larger scales at later stages. \\

\begin{figure}[h]
    \centering
    \includegraphics[width=8.5cm]{ 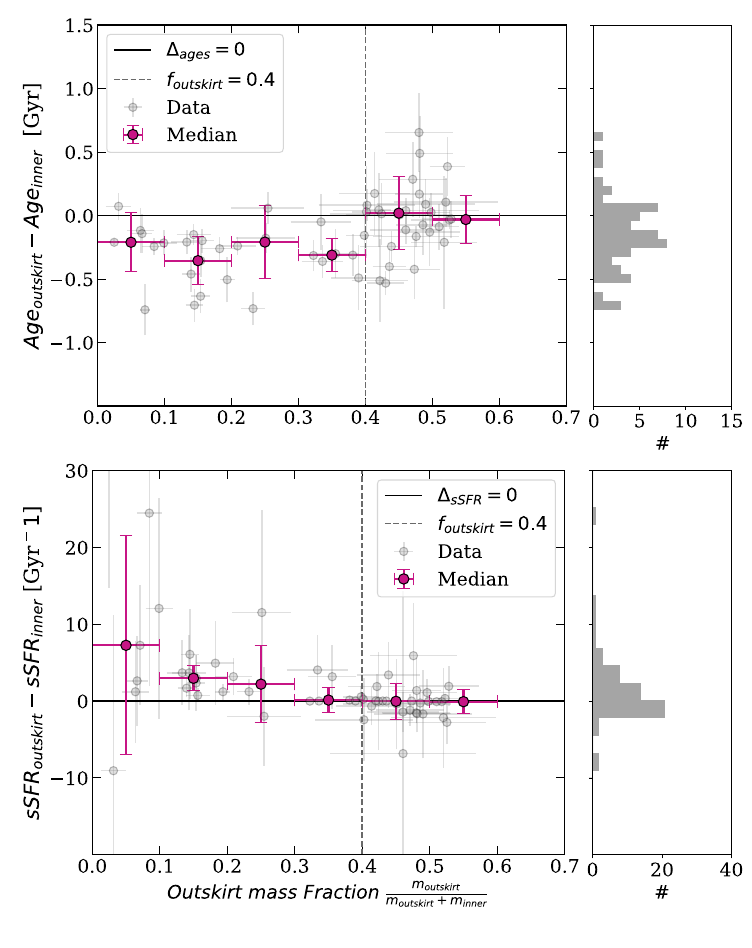}
    \caption{ Differences of Ages and sSFR between the outskirt and inner part as a function of the outskirt mass fraction $f_{outskirt}$. The data are shown in grey with the median profile in pink. For each distribution, the black solid line refers to no radial changes between the inner/outksirt part. }
    \label{fig:deltas}
\end{figure}

Because we are studying LRDs successor candidates over a long timescale of approximately 2 Gyr (2.5<z<7.5), all those scenarios could occur. To distinguish which one is more likely, we analyze in Figure \ref{fig:deltas} ages and specific star formation rates (sSFR) of the outskirt relative to the inner part as a function of the outskirt mass fraction. Two regimes appear to emerge, depending on whether $f_{\text{outskirt}}$ is below or above 0.4.

The first regime, where $f_{\text{outskirt}} < 0.4$, 90\% of galaxies exhibit a younger outskirt, favors scenarios (a) and (c). Additionally, also 90\% of these galaxies present a higher sSFR in the outer region, with a significant fraction (20\%) having $\text{sSFR}_{\text{outskirt}} - \text{sSFR}_{\text{inner}} > 5 \ \text{Gyr}^{-1}$. This excludes the merger scenario (b) when $f_{\text{outskirt}} < 0.4$, in which a flat specific star formation rate (sSFR) profile is more likely to be observed with dry mergers ((b).2), whereas a bluer inner part is more expected with wet mergers ((b).1), with a higher sSFR in the center, which is not observed in our sample. This might be a possible evolutionary path for LRDs, but our sample was not specifically constructed to trace this population. Moreover, \citet{Escudero_lonely_LRDs} found that the environment of LRDs tends to be less dense compared to other galaxies at similar redshifts, making the merger-driven evolutionary pathway less plausible. Finally, galaxies in this regime are very close to an LRD state, having a young, low-mass outskirt with the V-shape. Therefore, all the characteristics we would expect to observe in connection with feedback from AGN or supernovae in LRDs are most likely to be observed in this phase.

In the second regime, when $f_{\text{outskirt}} > 0.4$, the proportion of galaxies with a younger outskirt drops to 50\%, the ages are more homogeneous and the v-shape is no longer observed, which disfavors scenario (a) and (b).1. The sSFR profile shows the same characteristics, suggesting a homogeneous spatial repartition of both those properties, which favors the dry merger scenario of growth (b).2. Nevertheless, if the accretion rate declines with time (as it scales with $(1+z)^{5/2}$; \citealt{Dekel_2013}), the sSFR in the periphery would also decrease, potentially reaching values comparable to those of the inner region, thus making scenario (c) a plausible evolutionary pathway in this regime too. \\

If we now consider the outskirt mass fraction as a proxy for time, as suggested by Figure \ref{fig:outskirtgrowth}, direct accretion from filaments (scenario (c)) or AGN/supernova feedback (scenario (a)) could dominate initially even though the latter mechanism remains more uncertain (no strong evidence of gas reservoir except highly ionized gas), with the V-shape in place up to $f_{\text{outskirt}} = 0.3$, from $z = 6.5$ to $z = 4.5$, corresponding to approximately 1 Gyr. In this regime, galaxies have a young, low-mass outskirt with a higher sSFR. Over time, the accretion rate declines ($\propto(1+z)^{\frac{5}{2}}$; \citealt{Dekel_2013}), the sSFR and ages of the periphery respect to the inner region tend therefore to homogenies, as seen in our sample when \( f_{\text{outskirt}} > 0.4 \) in Figure \ref{fig:deltas}. Merger events could also occur, with a rate of $\approx 2 \text{ Gyr}^{-1}$ at  $z=4$ \citep{Mundy_merger_rate, Puskas_merger_rate}. Through kinematics, a redistribution of stars can take place, where age and sSFR differences should no longer be observable. Because the merger mechanism alone seems insufficient \citep{Inoyashi2025_LRDs} and even unlikely \citep{Escudero_lonely_LRDs} to explain the decline in the number density of LRDs, another mechanism is needed that could be either scenario (a) or (c). This mechanism could first be a phase of expansion due to accretion (scenario (c)) or the central AGN activity (scenario (a)), followed by a gradual decline in accretion (scenario (c)) or dry mergers (scenario (b)). At redshift \( 3.5 < z < 4.5 \), the stellar mass within the outskirts is \( \approx 10^9 \, M_\odot \), consistent with the estimated accreted mass around LRDs in \citet{chen_host_2024}. They investigated the origin of the extended emission observed in three different LRDs, attributing it to gas accretion. Then, they estimated accretion rates and the resulting stellar mass, assuming a star formation efficiency of 0.1. From redshift \( z = 6 \) to \( z = 4 \), this would yield a stellar mass of about \( 10^9 \, M_\odot \), strongly supporting the evolutionary scenario via accretion (scenario (c)).

\section{Conclusion}\label{sec:conclusion}

To date, the nature, origin, and evolutionary path of LRDs remain poorly understood. Their steep decline in number density at \( z < 4 \) suggests the action of physical processes that alter their observable properties, making them increasingly difficult to identify as they evolve. In this study, we investigated the possible fate of this population by identifying a set of candidate galaxies that may represent their descendants: galaxies as red as LRDs with a blue star-forming periphery at $z_\text{med} = 3.6$. \\

The population we identified appears to meet the key conditions expected of LRD descendants. They are approximately 1 kpc below the mass-size relation, exhibit stellar masses ($M_*= 10^{10.1\pm0.4} \ M_\odot$) and central stellar densities ($\Sigma_* \approx 10^{11.6 \pm1.9} \ M_\odot \ \text{kpc}^{-2}$) similar to those of LRDs. Their number density at $z=3\pm0.5$ is about the same as that of the LRDs at $5<z<7$, $ 10^{-4.15}\text{Mpc}^{-3}$ and they present the same rest-frame red colors. The blue rest-frame color is also consistent for sources at redshifts $z > 4$. However, approximately $50\%$ of the sources at redshifts $z < 4$ fall outside the blue color selection, indicating a redshift-dependent evolution.
 \\

Our interpretation is also strengthened by the finding that the fraction of mass contained in the outskirt increases with cosmic time, suggesting that we see the growth of outskirts of the LRDs taking place through cosmic time. The disappearance of the LRDs at lower z might therefore be accompanied by the acquisition of a stellar extension, which is compatible with a recently built extension. The growth scenario is strongly supported by the fact that the high-redshift sources in our sample are unresolved, with an upper limit on the effective radius of \(R_e \approx 260\) pc and an outskirt mass fraction of \(f_{\text{outskirt}} \approx 0.1\) at \(z \approx 4.5\). As redshift decreases, both \(f_{\text{outskirt}}\) and the effective radius \(R_e\) increase, with \(R_e \approx 285\) pc and \(f_{\text{outskirt}} = 0.3\) at \(z \approx 3.5\), and \(R_e \approx 590\) pc and \(f_{\text{outskirt}} = 0.5\) at \(z \approx 3\). Additionally, six LRDs previously analyzed in \citet{kocevski_rise_2024} are found at \( z > 5 \) within our sample. All of them exhibit rest-frame UV emission in their outer regions.\\

Our results suggest that the decline in the number density of LRDs is driven by the acquisition of a stellar component. As this component grows, the characteristic V-shaped SED fades, and the physical size of the galaxies increases. Consequently, both the V-shaped SED and compactness criteria used to select LRDs may no longer effectively identify this population.  \\

The growth of the stellar component can be explained by several mechanisms. However, accretion through cold streams or diffuse inflows is the more favored scenario, as indicated by the presence of younger outskirts when the galaxies are near the LRD phase, i.e., at \(f_{\text{outskirt}} < 0.4\). When \(f_{\text{outskirt}} > 0.4\), radial properties become more homogeneous, suggesting either a transition to a less significant accretion-driven growth phase or a growth phase dominated by dry merger events. \\

The galaxies analysed in this study exhibit extremely dense cores that emit predominantly in the NIR, along with non-negligible rest-frame UV emission tracing their extended structures. An intriguing avenue for further investigation is the identification of potential analogues at lower redshifts, around \( z \approx 2 \). \citet{Benton_bulge_dominated_JWST} report the existence of star-forming galaxies (SFGs) at \( z \approx 2 \) with high central stellar surface densities, suggesting that dense cores can already be in place in SFGs, similar to those found in quiescent galaxies (QGs). The most similar sources in our sample are those with high outskirt mass fractions (\( f_{\text{outskirt}} > 0.4 \)) and lower redshifts (\( z < 3 \)). These galaxies may represent systems evolving into bulge-dominated SFGs, given their already established compact, dense cores and  star-forming outskirts.
 \\

Further investigation is needed to constraint any AGN counterpart through spectroscopy in our sample. An Integral Field Unit (IFU) would also provide resolved metallically measurements, offering accurate proxies for stellar ages between the inner part and the outskirt. The study of dust with ALMA as a function of the outskirt mass fraction should also bring some insight into the buildup of dust in this population.  \\

\begin{acknowledgements}
      This work is based on observations made with the NASA/ESA/CSA 
      \emph{James Webb Space Telescope (JWST)}. The data were obtained from the Mikulski Archive for Space Telescopes at the Space Telescope Science Institute, which is operated by the Association of Universities for Research in Astronomy, Inc., under NASA contract NAS 5-03127 for JWST. These observations are associated with program \#1345. The CEERS NIRCam Images used in this article are available on the Mikulski Archive for Space Telescopes (MAST) at the Space Telescope Science Institute via doi: 
      \href{https://archive.stsci.edu/doi/resolve/resolve.html?doi=10.17909/z7p0-8481}{{10.17909/z7p0-8481}}.
      This project has received funding from the European Union’s Horizon 2020 research and innovation programme under the Marie Skłodowska-Curie grant agreement No 101148925. We would acknowledge the following open source software used in the analysis: \texttt{Astropy} \citep{Astropy} , \texttt{photutils} \citep{photutils}, \texttt{Numpy} \citep{numpy}.
   
\end{acknowledgements}

\bibliographystyle{aa}
\bibliography{LRDs}

\begin{appendix} 
\section{Outskirt detectability and surface brightness dimming}\label{sec:appendix}

\begin{figure}[h]
\includegraphics[width=7.8cm]{ 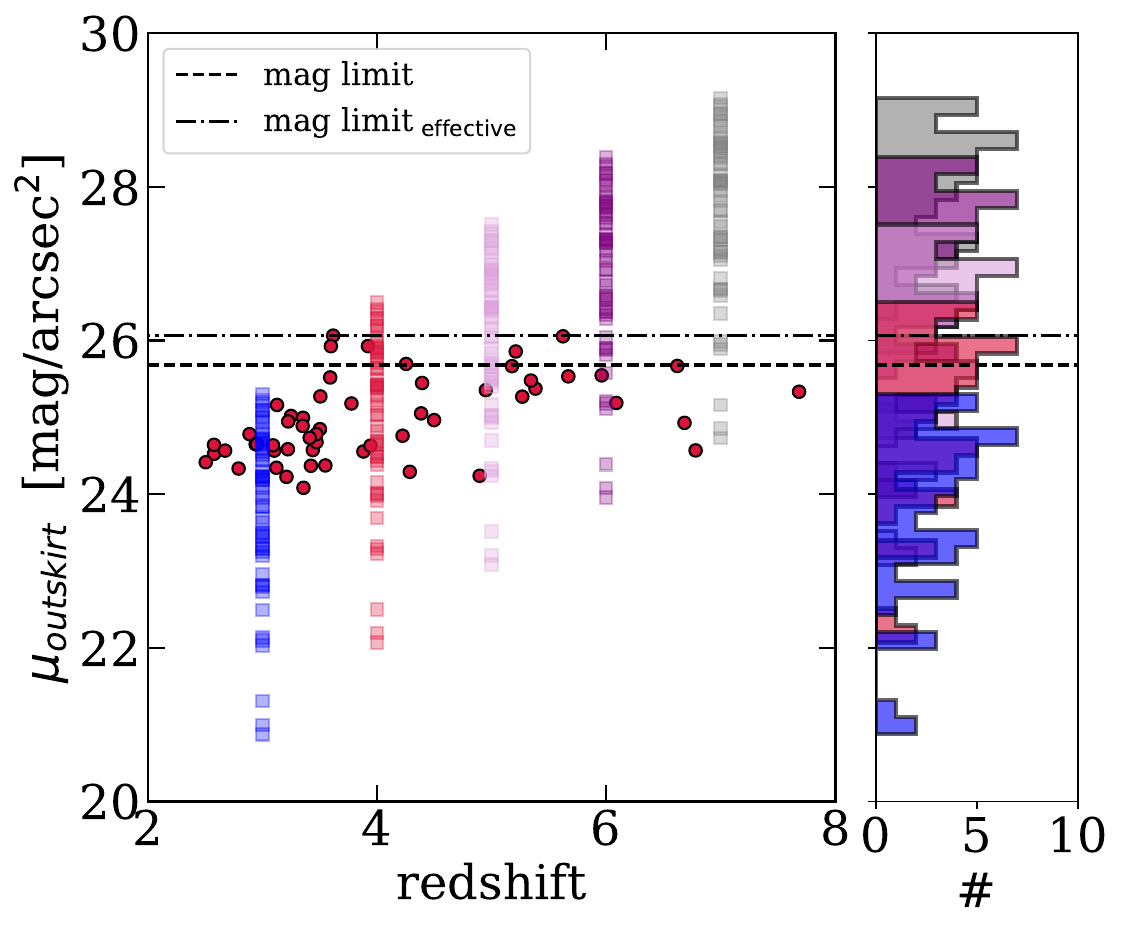}
\caption{Surface brightness density as a function of redshift for the outskirts in our sample (red dots), along with the observed surface brightness shifted to different redshifts (squares), including the dimming effect of the universe. The magnitude detection limit is shown as a dashed line, while the effective magnitude limit is indicated by a dash-dotted line. }
\label{appendix_1}
\end{figure}

In in this work, we select objects with a young, star forming periphery. Therefore, it would be interesting to investigate wether this outskirts component remains detectable or not due to the surface brightness dimming in the universe. To this end, we first compute the outskirts surface brightness density by measuring the surface of each outskirt (number of pixels) and using the observed magnitude in the F356W filter, which is both one of the reddest (hence better tracing stellar mass) and deepest bands available (with a 5$\sigma$ point source detection limit of 29.17 mag). The magnitude limit is shown as a dotted line on Figure \ref{appendix_1} and estimated using $\mu_{limit} = 29.17 + 2.5 \text{log}_{10}(2 \pi \text{HWHM}^2)$, with HWHM=0.08". The effective magnitude limit is also visible as a dashes-dotted line, corresponding to the faintest outskirt in our sample. As the observed flux decreases with redshift following \( I_\text{obs} \propto I_\text{0} (1+z)^{-4} \), $I_0$ being the emitted flux, the apparent magnitude increases. Denoting by \( m_1 \) the magnitude of a source at redshift \( z_1 \), its magnitude at a different redshift \( z_2 \) can be expressed as: \[m_2 = m_1 + 10 \log_{10} \left( \frac{1+z_1}{1+z_2} \right).\]
We then compute the surface brightness \( \mu_2 \) at redshift \( z_2 \) as: \[\mu_2 = m_2 + 2.5 \log_{10} (S_2)\]
where \( S_2 \) is the projected area of the outskirt in arcseconds squared at redshift \( z_2 \). Using this, we are able to shift our sample through redshift as visible on Figure \ref{appendix_1} from $z=3$ to $z=7$ and evaluate its detectability. As shown, outskirts remain above the detection limit (i.e., are detectable) for redshifts \( z < 4 \), indicating that the entire sample is observable in this range. However, at \( z = 5 \), approximately 50\% of the outskirts fall below the detection threshold, and this fraction increases to 80\% at \( z = 6 \). This implies that at \( z > 5 \), we are primarily sensitive to the most massive and extended outskirts. Consequently, a significant fraction of the population is likely missed, leading to an underestimation of the true number of such objects at high redshift.

\end{appendix}

\end{document}